\ifdefined\anonymous
\documentclass[sigconf,review,anonymous]{acmart}
\else
\documentclass[acmsmall,screen,nonacm]{acmart}
\fi

\usepackage[T1]{fontenc}
\usepackage{listings}
\usepackage{textcomp}
\usepackage{svg}
\usepackage{float}
\usepackage{todonotes}
\usepackage{hyperref}
\usepackage[capitalise,nameinlink,noabbrev]{cleveref}
\usepackage{colortbl}
\usepackage[inline]{enumitem}
\usepackage[xcolor]{changebar}

\presetkeys{todonotes}{inline}{}

\definecolor{light-gray}{gray}{0.85}
\definecolor{rltred}{rgb}{0.5,0,0}
\definecolor{rltgreen}{rgb}{0,0.5,0}
\definecolor{rltblue}{rgb}{0,0,0.5}

\hypersetup{
colorlinks=true,
urlcolor=rltblue, %
linkcolor=rltred,
citecolor=rltgreen,
}

\makeatletter
\def\lst@visiblespace{$\color{White}\bullet$}
\makeatother

\newcommand{\changebarcode}{\makebox[0pt][l]{\color{blue}\rule[-0.25em]{0.3mm}{1em}}}

\ifdefined\anonymous
\def\tool{\mbox{\textsc{BinaryFuzzer}}}
\def\toolURL{(blinded)}
\else
\def\tool{\mbox{\textsc{FormatFuzzer}}}
\def\toolURL{\texttt{\url{https://uds-se.github.io/FormatFuzzer/}}.}
\fi

\AtBeginDocument{%
  \providecommand\BibTeX{{%
    \normalfont B\kern-0.5em{\scshape i\kern-0.25em b}\kern-0.8em\TeX}}}

\setcopyright{none} %
\copyrightyear{2022}
\acmYear{2022}

\acmJournal{TOSEM}
\acmVolume{30}
\acmNumber{4}
\acmArticle{111}
\acmMonth{8}

\begin{document}

\title{\tool{}: Effective Fuzzing of Binary File Formats}

\author{Rafael Dutra}
\email{rafael.dutra@cispa.de}
\affiliation{%
  \institution{CISPA Helmholtz Center for Information Security}
  \streetaddress{Stuhlsatzenhaus~5}
  \city{Saarbrücken}
  \country{Germany}
  \postcode{66111}
}

\author{Rahul Gopinath}
\email{rahul.gopinath@cispa.de}
\affiliation{%
  \institution{CISPA Helmholtz Center for Information Security}
  \streetaddress{Stuhlsatzenhaus~5}
  \city{Saarbrücken}
  \country{Germany}
  \postcode{66111}
}

\author{Andreas Zeller}
\email{zeller@cispa.de}
\affiliation{%
  \institution{CISPA Helmholtz Center for Information Security}
  \streetaddress{Stuhlsatzenhaus~5}
  \city{Saarbrücken}
  \country{Germany}
  \postcode{66111}
}

\theoremstyle{theorem}
\newtheorem{tkaway}{Takeaway}
\def\takeaway#1{\begin{tkaway}#1\end{tkaway}}

\begin{abstract}
Effective fuzzing of programs that process structured binary inputs, such as multimedia files, is a challenging task, since those programs expect a very specific input format.
Existing fuzzers, however, are mostly \emph{format-agnostic,} which makes them versatile, but also ineffective when a specific format is required.

We present \tool{}, a generator for \emph{format-specific fuzzers}. \tool{} takes as input a \emph{binary template} (a format specification used by the 010~Editor) and compiles it into C++ code that acts as \emph{parser,} \emph{mutator,} and highly efficient \emph{generator} of inputs conforming to the rules of the language.

The resulting format-specific fuzzer can be used as a standalone producer or mutator in black-box settings, where no guidance from the program is available.
In addition, by providing mutable \emph{decision seeds,} it can be easily \emph{integrated} with arbitrary \emph{format-agnostic fuzzers} such as AFL to make them format-aware.
In our evaluation on complex formats such as MP4 or ZIP, \tool{} showed to be a highly effective producer of valid inputs that also detected previously unknown memory errors in \texttt{ffmpeg} and \texttt{timidity}.
\end{abstract}



\maketitle

\section{Introduction}
\label{sec:intro}

\begin{figure}[t]
  \center
  \includegraphics[width=10cm]{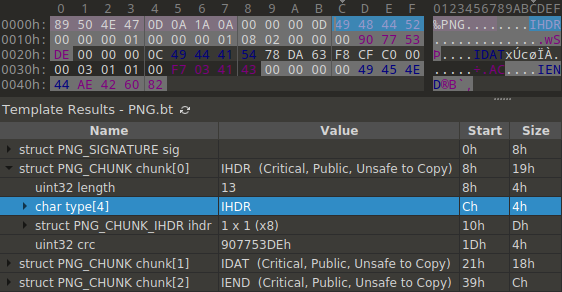}
  \caption{010 Editor displaying a PNG file.}
  \label{fig:010}
\end{figure}

Feedback-directed fuzzing tools such as AFL~\cite{afl} and libFuzzer~\cite{serebryany2016continuous} have shown great success in finding bugs and vulnerabilities in widely used software.
Such tools are generally \emph{format-agnostic}, meaning that they assume no knowledge about the input format when producing or mutating inputs.
This makes them easy to set up and deploy.
However, it also limits their usage to settings in which
\begin{enumerate*}
\item Sample input files to be mutated are available;
\item Instrumentation and feedback from the program under test is available;
\item The cost of producing myriads of invalid inputs is bearable.
\end{enumerate*}

The alternative to a format-agnostic fuzzer is to create a \emph{format-specific} fuzzer, making use of format knowledge to produce valid inputs.
However, creating such a fuzzer comes with considerable effort, in particular when one wants to reuse the guidance and analysis capabilities of feedback-directed fuzzers.

In this paper, we present a novel approach that combines the flexibility of format-agnostic fuzzers with the efficacy of format-specific fuzzers.
Our \tool{} framework leverages existing \emph{binary templates}---specifications for input formats, from JPEG to PCAP---to
produce or mutate valid file inputs, even in black-box settings where no guidance from the program is available;
and integrate with arbitrary \emph{format-agnostic fuzzers} such as AFL to make them format aware.
In all settings, using \tool{} \emph{increases coverage in the program under test,} covering code not reached by a format-agnostic fuzzer, and thus increasing the chance of detecting bugs and vulnerabilities.%

How does \tool{} work? The inputs for \tool{} are \emph{existing input specifications.}
The 010~Editor~\cite{010editor} is a commercial editor which allows users to view and edit binary files with detailed structural information, as shown in \Cref{fig:010}.
To parse its inputs, the editor relies on human written specification files, known as \emph{binary templates}~\cite{010editortemplates} to specify the different sections of a file.
Those binary templates are written in a C-like language and have been community-developed for over ten years, resulting in a repository of more than 200~different specifications for popular binary file formats~\cite{010editortemplatesrepository}.
So, for many popular fuzzing targets, a binary template which describes its input format is already available online.
For instance, the majority of bugs listed under AFL's bug trophy case~\cite{afl} come from programs which process popular binary formats for which there is a public specification.
Of special interest are multimedia file formats, such as images, audio or video, which often come from untrusted sources, and therefore represent a large attack surface, with potentially serious security consequences.

\begin{lstlisting}[caption={PNG binary template (extract).},label={lst:pixel},float=t]
typedef struct {
    byte btRed;
    byte btGreen;
    byte btBlue;
} PNG_PALETTE_PIXEL;

struct PNG_CHUNK_PLTE (int32 chunkLen) {
    PNG_PALETTE_PIXEL plteChunkData[chunkLen/3];
};
\end{lstlisting}

As an example of a binary template, consider \Cref{lst:pixel}, showing an excerpt of the original template for PNG files.
We see that (like a C program), it defines \emph{types} such as \texttt{PNG\_PALETTE\_PIXEL} consisting of three byte values. The following \texttt{struct} fragment defines a chunk of such pixels named \texttt{PNG\_CHUNK\_PLTE}, whose overall (parameterized) length is \texttt{chunkLen}.
The full PNG format contains dozens of such chunk definitions, all formalized within the PNG binary template; these also include executable code that computes context-sensitive information such as checksums (whose inference is a major roadblock for format-agnostic fuzzers).

\tool{} takes as input one of these binary templates---say, a PNG template. It then \emph{compiles} it into:
\begin{enumerate}
\item A highly efficient \emph{generator} that produces outputs in the format specified (i.e., a PNG fuzzer); and
\item A \emph{parser} that reads in existing inputs in the format specified (i.e., a PNG parser).
\end{enumerate}
The \emph{parser} can immediately be used in existing \emph{parser-enhanced} fuzzers such as AFLSmart~\cite{pham2019smart}, effectively giving us a format-specific fuzzer that evolves, say, given PNG seed inputs; this can be repeated with hundreds of different templates, producing hundreds of format-specific, coverage-guided fuzzers.
The \emph{generator,} on the other hand, can be used as a \emph{standalone} generator of inputs that all conform to the format specifications.
This is useful in black-box settings, where feedback-driven fuzzers cannot be applied.
Furthermore, the combination of \emph{parser} and \emph{generator} allows \tool{} to \emph{mutate} existing inputs in a way that ensures their validity.
Finally, \tool{} allows to \emph{integrate format-agnostic fuzzers and make them format-aware,} either by having them apply smart (format-aware) mutations on the inputs, or by having them mutate \emph{decision seeds}---strings of bytes that represent decisions taken during parsing and generating.
Both integrations bring together the best of format-agnostic guidance and format-aware mutations, and increase the reach of format-agnostic fuzzers.

\newcommand*\step[1]{\tikz[baseline=(char.base)]{
    \node[shape=circle,draw,inner sep=1pt] (char) {\footnotesize #1};}}
\renewcommand\step[1]{(#1)}
\newcommand{\sampleFormat}{{\boldmath$F$}}

\begin{figure}[t]
\includegraphics[width=10cm]{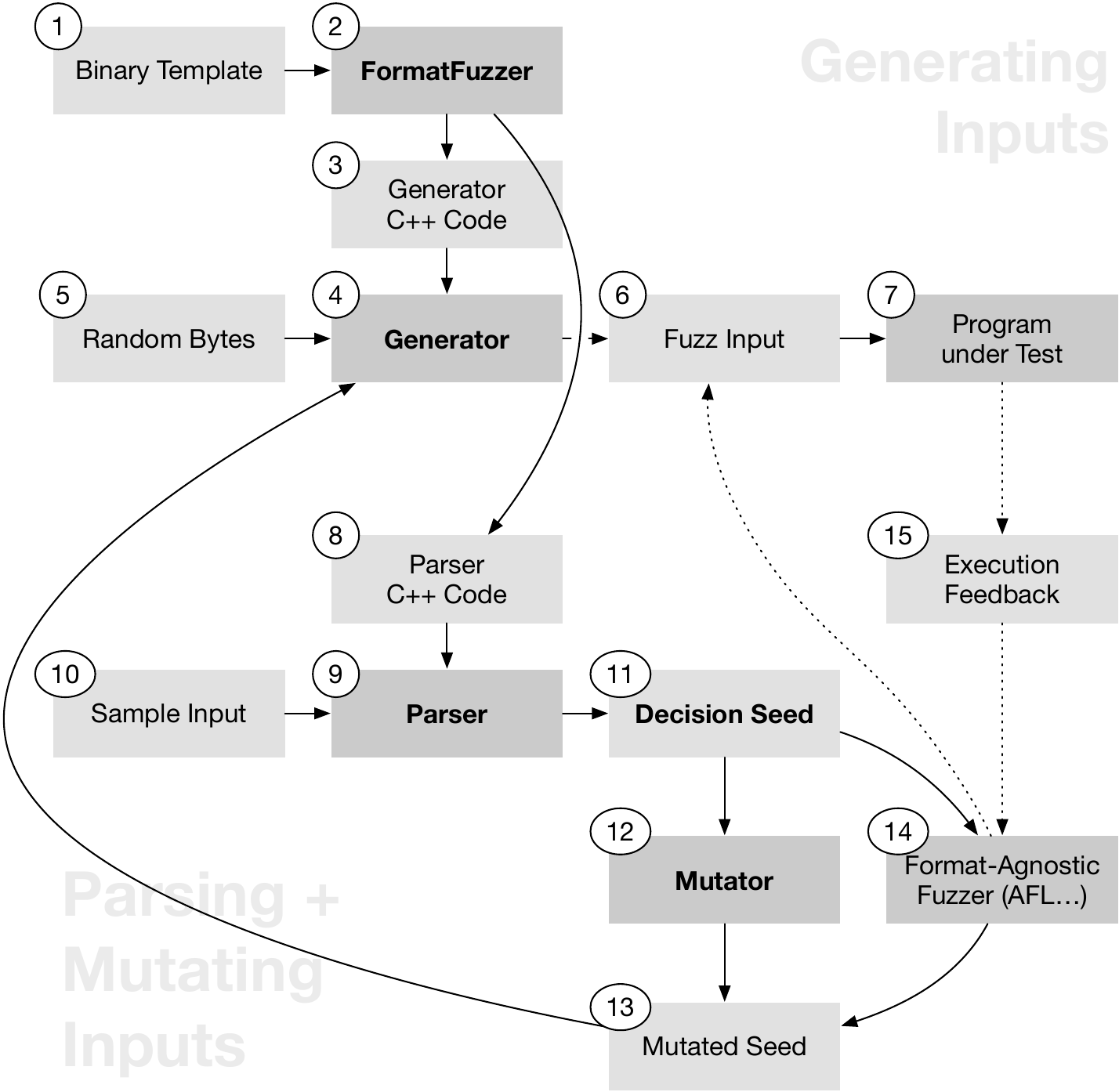}
\caption[\tool{} overview]{\tool{} overview.
From a \emph{binary template}~\step{1} that specifies an input format~\sampleFormat, \tool{}~\step{2} generates C++ code~\step{3} that compiles into a highly efficient \emph{generator}~\step{4} for~\sampleFormat.
In a \emph{black-box testing} setting, the generator uses a source of random bytes~\step{5} to generate fuzz input~\step{6} in the format~\sampleFormat{} for the program under test~\step{7}. From the same binary template, \tool{} also generates C++ code~\step{8} for a \emph{parser}~\step{9} for~\sampleFormat.
In a \emph{mutation} setting,  this parser transforms a sample input~\step{10} into a \emph{decision seed}~\step{11}, which encodes the decisions made while parsing the sample into a string of bytes.
\tool{} can \emph{mutate} this seed~\step{12} (for example, replacing one chunk); feeding the mutated seed~\step{13} into the generator yields a mutated fuzz input still conforming to~\sampleFormat.
Finally, for an \emph{evolutionary} setting, the mutator can also be a format-agnostic fuzzer~\step{14} such as AFL, which then mutates and evolves seed strings based on execution feedback~\step{15}.
Such a fuzzer can also apply out-of-format mutations to the fuzz input directly.}
\label{fig:overview}
\end{figure}

Nevertheless, some manual effort is still required.
The hundreds of existing binary templates work well for \emph{parsing;} but to make effective \emph{generators,} one has to further refine them.
These refinements, however, are easy to write and the feature-full language of binary templates is rich enough to express virtually all needed features of binary file formats, including length fields, checksums, complex structures and constraints between variables.
Extending a binary template and compiling it is a one-time effort for each format; and once one has, say, a full-fledged PNG specification, one gets a PNG fuzzer, a PNG parser, a PNG mutator, and a variant of your favorite format-agnostic, feedback-directed fuzzer that produces thousands of valid PNG files per second---with only a fraction of the effort it would take to build such a fuzzer from scratch.

In the remainder of the paper, we detail the techniques behind \tool{} and how they improve the state of the art in fuzzing. Specifically, we make the following contributions:

\begin{enumerate}
\item \textbf{A novel description format for generating binary inputs.}
There are hundreds of well-curated \emph{binary templates} available that describe just as many binary file formats.
In \Cref{sec:bt}, we describe their structure; and in \Cref{sec:btgen} we extend them with new features for \emph{generating} valid inputs.

\item \textbf{Decision seeds to synchronize generators and parsers.}
\label{item:decision}
\tool{} takes binary templates to produce C++ code (\Cref{fig:overview}) that compiles (\Cref{sec:compiler}) into
specialized and highly efficient generators and parsers, including several context-sensitive features (\Cref{sec:btgen}).
Generator and parser synchronize over \emph{decision seeds}---a sequence of bytes reflecting decisions taken during generating or parsing (\Cref{sec:decisions}). After parsing an input, the resulting decision seed allows the generator to re-take the same decisions (but now for generating), which reproduces the input exactly.
As far as we know, \tool{} is the first framework to enable this bidirectional transformation between decision seeds and binary files.
This allows us to apply novel smart mutations (\Cref{sec:mutating}) over decision seeds which are obtained from a high-quality corpus of binary files.

\item \textbf{Making format-agnostic fuzzers input-format aware.}
Mutating bytes in the decision seed allows to explore the entire domain of inputs.
If a format-agnostic fuzzer is thus set to mutate and evolve decision seeds, \tool{} can then translate any such mutation into a valid binary input.
This enables a generator-based approach for integrating \tool{} with any format-agnostic fuzzer, leveraging its mutation and evolution strategies (\Cref{sec:strategies}).
In addition, \tool{} provides novel smart mutations, which operate over decision seeds and can therefore respect contextual information in mutated files.
This enables another highly effective approach to improve any format-agnostic fuzzer, by leveraging the format-specific \emph{smart mutations} provided by \tool{} (\Cref{sec:strategies}).
\item \textbf{A detailed evaluation of all \tool{} elements.}
We evaluated \tool{} (\Cref{sec:evaluation}) on 10 different programs and file formats (PNG, JPEG, GIF, MIDI, MP4, ZIP, PCAP, AVI, WAV, BMP) and found that:
\begin{enumerate}
\item \tool{} very quickly produces valid inputs, as standalone generator as well as mutator.
\item \tool{} integrates well with format-agnostic fuzzers such as AFL, reaching lines otherwise not covered. In our initial fuzzing experiments, \tool{} detected a number of previously unknown (and potentially exploitable) segmentation faults and aborts in \texttt{ffmpeg} and \texttt{timidity}.
\end{enumerate}
\end{enumerate}
\tool{} is available as open source and enjoys a quickly growing user community. For details, see \texttt{\url{https://github.com/uds-se/FormatFuzzer}}.

\section{The Binary Template Language}
\label{sec:bt}

We start our exploration into \tool{} with a description of its central input---\emph{binary templates.}
The \emph{binary template language}~\cite{010editortemplatelanguage} was designed to fully specify any binary format.
Its syntax and semantics are close to the C programming language, the main difference being that it describes not programs, but data formats. Its main elements are:

\begin{description}
  \item[Variable (input data) declarations.] Every variable declared in the binary template is \emph{directly mapped to a sequence of bytes in the input.}
  If a template starts with \texttt{uint32 len; char s[len]; int64 n;}, this means that the input consists of
  \begin{enumerate}
    \item a 32-bit unsigned integer (\texttt{len}) in the first four bytes;
    \item an array of characters of length \texttt{len} in the next \texttt{len} bytes; and
    \item a 64-bit integer (\texttt{n}) in the next four bytes.
  \end{enumerate}
  Note that array sizes can be any expression including variables and functions; this allows to specify variable lengths.

  \item[Local variables.]
  When a variable is defined with the keyword \texttt{local}, it is \emph{not} mapped to input contents, but acts like a traditional variable stored in memory.
  It can be freely used in computations and control-flow decisions by the binary template.

  \item[Type declarations.] As in C, one can define \emph{types} using the \texttt{typedef} and \texttt{struct} constructs. \Cref{lst:pixel} defines a type \texttt{PNG\_PALETTE\_PIXEL}, composed of three individual bytes.

  Besides native types (such as \texttt{uint32} and \texttt{string}) and \texttt{struct} types, variables can also have \texttt{enum} types, where the list of possible values is explicitly provided.

  \item[Parameters.] Unlike C, type and variable declarations can take \emph{parameters.} The definition of \texttt{PNG\_CHUNK\_PLTE} in \Cref{lst:pixel}, for instance, is parameterized with a \texttt{chunkLen} parameter, which sets the size of the contained array.

  \item[Conditionals.] Binary templates incorporate all the usual control-flow constructs (\texttt{if}, \texttt{else}, \texttt{while}, \texttt{switch}, etc.); their bodies would again contain variable declarations.
  These are used to express \emph{alternatives} and \emph{loops} in the input.
  For instance, to express a sequence of chunks, whose actual type depends on a header, we could use a loop and conditions as in \Cref{lst:loops}.

  \begin{lstlisting}[caption={Loops and alternatives in a binary template.},label={lst:loops},float=h]
while (!FEof()) {
    uint32  length;
    char   type[4];
    if (type == "IHDR")
        PNG_CHUNK_IHDR    ihdr;
    else if (type == "PLTE")
        PNG_CHUNK_PLTE    plte(length);
    /* ... */
    if (type == "IEND")
        break;  /* break out of while loop */
}
  \end{lstlisting}

  \item[Built-in functions.]
  The \texttt{while} condition in \Cref{lst:loops} invokes a \emph{function} named \texttt{FEof()} to check for an end-of-file condition.
  The language provides several such functions to allow for greater control during parsing, making the language expressive enough to support complex constructs from hundreds of binary formats.
  Examples include \texttt{FTell()}, which returns the current file position, \texttt{FSeek()}, which allows jumping to a different file position for non-sequential parsing, and \texttt{FileSize()}, which returns the file size.
  Advanced lookahead functions such as \texttt{FindFirst()} and \texttt{FindAll()} allow searching the file for specific tokens.

  Lookahead functions allow the parser to \emph{peek} at a few later bytes in the file without advancing the file position, and can be used for control-flow decisions. %
  For instance, peeking at the first bytes of a \texttt{struct} ahead of time can help determine which \texttt{struct} type should be parsed.
  Examples of lookahead functions are \texttt{ReadByte()} and \texttt{ReadBytes()}.

  Besides I/O functions, a multitude of string and math functions is available, including conversion and checksum algorithms.

  \item[Custom functions.]
  The language also allows defining own functions, which follow C function syntax and semantics, again allowing variable declarations for describing input data.
  As all the operators from C are also supported, this makes binary templates a \emph{Turing-complete description of input formats.}

  \item[Error handling.]
  In the case of irrecoverable parsing errors, the language also allows early termination, using a top-level \texttt{return} statement.
\end{description}
A full definition of syntax and semantics of the binary template language is available online~\cite{010editortemplatelanguage}.

We would like to note that our choice for binary templates as the base for \tool{} is purely pragmatic.
Binary templates have been shown to be able to precisely document hundreds of formats; they are reasonably well documented; and as we will show in this paper, they are not too hard to adapt for effective fuzzing.
Whether all the subtleties of input formats can be specified in a more elegant or useful way is yet to be proven.

\section{Binary Templates for Generation}
\label{sec:btgen}

In order to use binary templates to \emph{generate} new files from scratch, some decisions need to be made during the generation process.
Most notably, a value will have to be chosen for every variable that is declared, and also for every lookahead function that is called.
If such values were to be chosen uniformly at random from the entire range of possible values (for example, $2^{32}$ possible values for a \texttt{uint32}), most variables would end up having nonsensical values, rendering the file invalid.
Therefore, we extend the language of binary templates in a number of aspects:

\begin{description}
\item[Choices of valid values.]   Our first extension allows the specification of an \emph{array of valid values to choose from.} When a variable is declared, this array of choices can be assigned as an \emph{initialization list.}  \Cref{line:bits} of \Cref{lst:ihdr} tells the generator to pick a value for \texttt{bits} uniformly from five possible choices.
\item[Enumerations.]   To handle enumerations, the generator chooses a value uniformly from the set of the enumerated values (e.g., \texttt{PNG\_COLOR\_SPACE\_TYPE} in \Cref{lst:ihdr}), which is usually the appropriate behavior.
But it is also possible to specify a different set of choices through an initialization list.

\item[Lookaheads.] For lookahead functions, we extend the language to allow a new argument specifying the possible values to choose from.
One example is the array \texttt{colors} passed as an argument to the \texttt{ReadByte()} call in \Cref{line:ReadByte} of \Cref{lst:ihdr}.

\item[Allowing invalid choices.] Finally, we also allow the generator to deviate from the specified behavior and pick any value from the entire range of $2^8$ possible values for \texttt{ubyte} with a small probability of about 1\%, by performing an action that we call an ``evil'' decision.
Such evil decisions are enabled by default, but can be turned on and off during generation by calling a function \texttt{SetEvilBit()}, which takes as input the new value for the flag and returns the old value.
When enabled, they can give the fuzzer greater diversity of generated files by allowing the fuzzer to explore a small number of invalid choices during generation.
\end{description}

\begin{lstlisting}[caption={Contents of the IHDR chunk, where \texttt{bits} depends on \texttt{color\_type}. Changes made to enable generation are highlighted in \color{blue}blue\color{black}.},label={lst:ihdr},float=h]
typedef enum <byte> {
    GrayScale=0, TrueColor=2, Indexed=3, AlphaGrayScale=4, AlphaTrueColor=6
} PNG_COLOR_SPACE_TYPE;

typedef struct {
^\changebarcode^    uint32   width @<min=1, max=24>@;^\label{line:width}^
^\changebarcode^    uint32   height @<min=1, max=24>@;^\label{line:height}^
^\changebarcode^    @local byte colors[] = { GrayScale,@ @TrueColor, Indexed, /* ... */ };@
^\changebarcode^    @switch (ReadByte(FTell() + 1, colors)) {   /* color_type */@^\label{line:ReadByte}^
^\changebarcode^    @case GrayScale:@
^\changebarcode^        @ubyte    bits = { 1, 2, 4, 8, 16 };^\label{line:bits}^@
^\changebarcode^        @break;@
^\changebarcode^    @case TrueColor:@
^\changebarcode^        @ubyte    bits = { 8, 16 };^\label{line:truecolor}^@
^\changebarcode^        @break;@
^\changebarcode^    @case Indexed:@
^\changebarcode^        @ubyte    bits = { 1, 2, 4, 8 };@
^\changebarcode^        @break;@
^\changebarcode^    @/* ... */@
^\changebarcode^    @}@
    PNG_COLOR_SPACE_TYPE color_type;^\label{line:color_type}^
    PNG_COMPR_METHOD     compr_method;
    PNG_FILTER_METHOD    filter_method;
    PNG_INTERLACE_METHOD interlace_method;
} PNG_CHUNK_IHDR;
\end{lstlisting}

Let us now put these extensions into action and show what changes had to be made to the binary template in order to support the generation of valid inputs.
Our running example, PNG, is representative of the changes that are required for other file formats as well.
This section exhaustively describes all the changes which were needed to generate valid PNGs.
For any format we have seen so far, all its modifications also fall into the classes described here.
All our changes are highlighted in \color{blue}blue \color{black} in the code listings.

\subsection{Magic Values}
\label{sec:magic}

\begin{lstlisting}[caption={PNG signature: the magic bytes are mined automatically.},label={lst:sig},float=t]
typedef struct {
    uint16 btPngSignature[4];
} PNG_SIGNATURE;

^\changebarcode^ @local int evil = SetEvilBit(false);@
PNG_SIGNATURE sig;
^\changebarcode^ @SetEvilBit(evil);@
if (sig.btPngSignature[0] != 0x8950 ||
        sig.btPngSignature[1] != 0x4E47 ||
        sig.btPngSignature[2] != 0x0D0A ||
        sig.btPngSignature[3] != 0x1A0A) {
    Warning("File is not a PNG image.");
    return -1;
}
\end{lstlisting}

The simplest common feature of binary formats is the use of \emph{magic values}, where certain bytes in the file need to have a specific value.
For example, a PNG file needs to start with a signature containing the bytes \texttt{"\textbackslash x89504E470D0A1A0A"}.
In the original PNG template, there was already a check which compares the signature to the expected value and terminates the template if the signature is invalid, as shown in \Cref{lst:sig}.
In this case, \tool{} is able to automatically mine such comparisons (with operators \texttt{!=} or \texttt{==}) when analyzing the source code of the binary template and remember the magic values that should be used.
For example, the value \texttt{0x8950} will be remembered as a good value to use for index~\texttt{0} in the array \texttt{btPngSignature}.
Such comparisons can also be mined inside struct definitions and functions.

The only change made to the binary template for the signature generation was adding two calls to \texttt{SetEvilBit()}, which disable evil decisions temporarily for the signature generation, re-enabling them afterwards.
This essentially marks the generation of the \texttt{sig} variable as \emph{strict}, which is useful because any file with an incorrect PNG signature is not a valid PNG file and would be immediately rejected by the target program.

\subsection{Size Fields}
\label{sec:size}

Another common feature in binary formats is the presence of size fields, whose value correspond to the size of a certain structure in the file, or some particular position or index inside the file.
Such size fields are context-sensitive features, and binary formats using such features are
not expressible using context-free grammars.
It is particularly crucial that size fields are generated correctly, because a wrong value for a size field would completely change the interpretation of the remaining bytes of the file.

\begin{lstlisting}[caption={Definition of a PNG chunk, where we can see the changes to correct the chunk \texttt{length} and \texttt{crc}.},label={lst:chunk},float=t]
typedef struct {
^\changebarcode^    uint32  length @<min=1, max=16>@;^\label{line:length}^
    local int64 start = FTell();^\label{line:start}^
    char   type[4];
    if (type == "IHDR")
        PNG_CHUNK_IHDR    ihdr;
    else if (type == "PLTE")
        PNG_CHUNK_PLTE    plte(length);
    /* ... */
^\changebarcode^    else if (length > 0 @&& type != "IEND"@)^\label{line:IEND}^
        ubyte   data[length];^\label{line:arraydata}^
    local int64 end = FTell();
^\changebarcode^    @local uint32 correct = end - start - 4;@^\label{line:fixstart}^
^\changebarcode^    @if (length != correct) { /* Fix it */@
^\changebarcode^        @FSeek(start - 4);@
^\changebarcode^        @local int evil = SetEvilBit(false);@
^\changebarcode^        @uint32  length = { correct };@^\label{line:correct}^
^\changebarcode^        @SetEvilBit(evil);@
^\changebarcode^        @FSeek(end);@
^\changebarcode^    @}@^\label{line:fixend}^
    local uint32 crc_calc = Checksum(CHECKSUM_CRC32, start, end-start);
^\changebarcode^    uint32  crc @= { crc_calc }@;^\label{line:crc}^
    if (crc != crc_calc)
        Warning("Bad CRC %
} PNG_CHUNK;
\end{lstlisting}

In the PNG format, the chunk \texttt{length} is the first field in every chunk, as seen in \Cref{line:length} of \Cref{lst:chunk}.
There, we have extended the definition of the \texttt{length} variable with additional metadata \texttt{min=1} and \texttt{max=16} to restrict the choices of values for this variable to a small range.
This is important since this variable will be used as the length of an array, as seen in \Cref{line:arraydata} of \Cref{lst:chunk}.
If no such bounds are specified, \tool{} still samples values for integer variables from a skewed distribution that prioritizes small positive integers, but still allows arbitrary large or negative integers with a smaller probability.

\begin{lstlisting}[caption={Sequence numbers require a global counter.},label={lst:seq},float=h]
^\changebarcode^ @local uint32 seq_num = 0;@
struct PNG_CHUNK_FDAT {
^\changebarcode^    uint32 sequence_number @= { seq_num++ }@;
    ubyte  frame_data[length-4];^\label{line:frame_data}^
};
\end{lstlisting}

One challenge in generating PNG chunks which is also common in other file formats is that the \texttt{length} field appears before the data it refers to.
And it is not always possible to predict beforehand what the length of certain data should be before generating that data.
In addition, a bad choice for \texttt{length} could lead to the length of some array becoming too large or negative, such as when an integer underflow happens in \Cref{line:frame_data} of \Cref{lst:seq}.
The strategy we use to tackle those issues is to use the initial value sampled for \texttt{length} only as a hint.
So if, for example, this integer underflow happens when defining the array \texttt{frame\_data}, we ignore the value of \texttt{length} and choose some small positive integer to use as the length of the array.
This behavior can be enabled in the binary template with a call to the function \texttt{ChangeArrayLength()}.

Now, since the chosen value of \texttt{length} is used only as a hint, the generated data may end up being larger or smaller than expected.
This needs to be corrected. Otherwise the chunk will have a different length than what is specified in the \texttt{length} variable, leading to all following chunks being interpreted incorrectly.
Therefore, after generating the data inside a chunk, we check if the length is correct.
In case of a discrepancy, we go back and overwrite the value of \texttt{length} with the correct value, as shown in \Crefrange{line:fixstart}{line:fixend} of \Cref{lst:chunk}.
There, the \texttt{FSeek()} function is used to jump to a different position in the file.
Note that we temporarily disable evil decisions when fixing the value of \texttt{length}, because any wrong value would completely change the file parse tree.
Such common size fixes can also be simplified by the use of a macro \texttt{FIX\_VARIABLE()}, so that this whole block of code can be replaced by a call to the macro.
\begin{lstlisting}[numbers=none,showspaces=true]
// arguments: type ,  name ,  correct value , position
FIX_VARIABLE(uint32, length, end - start - 4, start - 4);
\end{lstlisting}

\subsection{Complex Constraints}

Next we discuss how to satisfy some more complex constraints between variables in the template.

\begin{description}
    \item[Relationship between variables.]
One common pattern happens when the value of one variable \texttt{x} restricts the possible values for another variable \texttt{y}.
If \texttt{x} is generated first, then we can simply use the value of \texttt{x} in control-flow decisions or initialization lists that will define how \texttt{y} is generated.
For example, a PNG chunk of type \texttt{"IEND"} must have length 0.
This can be ensured by using the value of \texttt{type} in a control-flow decision in \Cref{line:IEND} of \Cref{lst:chunk} to ensure no \texttt{data} will be generated if \texttt{type} is \texttt{"IEND"}.

In case \texttt{x} will only be generated after \texttt{y}, we can still look first at the value of \texttt{x} using a lookahead function.
\Cref{lst:ihdr} shows the implementation of the IHDR chunk, where the possible values for the \texttt{bits} variable are restricted by the value of the next variable \texttt{color\_type}.
By using the lookahead function \texttt{ReadByte()} in \Cref{line:ReadByte}, we can first look at the value at position \texttt{FTell()+1}, one byte after the current position.
This will be the value of \texttt{color\_type}, and can be used in a \texttt{switch} to determine the possible values for \texttt{bits}.

In generation mode, \texttt{ReadByte()} will use the array \texttt{colors} provided as an argument to choose which value should be generated at position \texttt{FTell()+1} in the file.
Once a value is chosen by the lookahead function, it is fixed and cannot be overwritten.
Therefore, when \texttt{color\_type} is declared later in \Cref{line:color_type}, it is guaranteed to have the same value that was previously chosen by \texttt{ReadByte()}.

\item[Checksums.]
Checksums are a common context-sensitive feature of binary formats.
Luckily, the binary template language already provides a helper function \texttt{Checksum()} which implements common checksum algorithms.
Therefore, checksums can be handled simply by computing the correct checksum value and specifying it as the desired value for the checksum variable, as shown in \Cref{line:crc} of \Cref{lst:chunk}.

\item[Global state.]
Some features, such as sequence numbers which are incremented every time, require some global state to be generated correctly.
This can be handled by defining a \texttt{local} variable of global scope, as in \Cref{lst:seq}.
\end{description}

\subsection{Chunk Ordering}
\label{sec:ordering}

A big challenge in generating valid files is choosing a valid sequence of chunk types.
The set of possible types for the next chunk is usually context-sensitive, depending on which chunks have already been generated.
In PNG, for instance, the chunks IHDR, IDAT and IEND are mandatory.
The first chunk must be IHDR and the last chunk must be IEND.
Some optional chunks, such as tIME, can appear before or after IDAT.
Other optional chunks, such as bKGD, can only appear before IDAT.

If we are only interested in parsing (not generating) PNG files, the following implementation suffices to parse a sequence of chunks.
\begin{lstlisting}[numbers=none]
while (!FEof())
    PNG_CHUNK chunk;
\end{lstlisting}
However, to generate a \emph{valid} sequence of PNG chunks, we need a way to specify the set of possible choices and allow those choices to change for each new generated chunk.

\begin{figure}
\begin{lstlisting}[caption={Use of \texttt{ReadBytes()} to define chunk ordering.},label={lst:ReadBytes}]
^\changebarcode^ @ChangeArrayLength();@
^\changebarcode^ @local char chunk_type[4];@
^\changebarcode^ @local string preferred[] = { "IHDR" };@^\label{line:pref1}^
^\changebarcode^ @local string possible[] = { "IHDR" };@^\label{line:pos1}^
^\changebarcode^ @while (ReadBytes(chunk_type, FTell() + 4, 4, preferred, possible, 0.25)) {@^\label{line:ReadBytes}^
    PNG_CHUNK chunk;^\label{line:chunk}^
^\changebarcode^    @switch (chunk_type) {@
^\changebarcode^    @case "IHDR":@
^\changebarcode^        @switch (chunk.ihdr.color_type) {@
^\changebarcode^        @case GrayScale:@
^\changebarcode^            @local string preferred[] = { "IDAT" };@^\label{line:pref2}^
^\changebarcode^            @local string possible[] = { "tIME", "tEXt", "pHYs", /*...*/ "bKGD", "IDAT" };@^\label{line:pos2}^
^\changebarcode^            @break;@
^\changebarcode^        @/* ... */@
^\changebarcode^        @}@
^\changebarcode^        @break;@
^\changebarcode^    @case "IDAT":@
^\changebarcode^        @local string preferred[] = { "IEND" };@^\label{line:pref3}^
^\changebarcode^        @possible -= ("IDAT", "pHYs", /*...*/ "bKGD");@^\label{line:pos3}^
^\changebarcode^        @possible += "IEND";@^\label{line:pos32}^
^\changebarcode^        @break;@
^\changebarcode^    @/* ... */@
^\changebarcode^    @case "IEND":@
^\changebarcode^        @local string preferred[0];@^\label{line:pref4}^
^\changebarcode^        @local string possible[0];@^\label{line:pos4}^
^\changebarcode^        @break;@
^\changebarcode^    @}@
^\changebarcode^ @}@
\end{lstlisting}
\end{figure}

To solve this problem, we have extended the lookahead function \texttt{ReadBytes()} with 3 new arguments: an array of preferred values, an array of possible values and a probability to pick from the preferred values.
We have also added new operators \texttt{-=} and \texttt{+=} which can add or remove values from such arrays.
\Cref{lst:ReadBytes} shows a binary template that can generate a sequence of PNG chunks satisfying the ordering constraints.
Here, function \texttt{ReadBytes()} in \Cref{line:ReadBytes} needs to choose the 4-byte value which will be written to position \texttt{FTell()+4} in the file, which is the location of the \texttt{type} array on the next chunk.
The chosen value is also returned in the \texttt{chunk\_type} variable, which is passed by reference to the function.

The \texttt{preferred} array is used to specify what is the next mandatory chunk, while the \texttt{possible} array includes all chunks which could be inserted in the next position.
With a probability of 0.25, \texttt{ReadBytes()} will try to pick a value from \texttt{preferred}, to ensure that it makes progress towards finishing the generation and does not spend too much time creating lots of optional chunks.
With the complementary probability of $1 - 0.25$, \texttt{ReadBytes()} will pick a value from the \texttt{possible} array (or even a completely random value if an evil decision is made).
If the \texttt{preferred} array is empty, \texttt{ReadBytes()} can choose a value from \texttt{possible}, but it is also allowed to not choose any value, not write any value to the file and return \texttt{false}, leaving the \texttt{while} loop.
This signals that the file is now complete and no more chunks will be added.

With no evil decisions, a valid chunk order is guaranteed.
The first chunk must be IHDR as the only option specified in \Cref{line:pref1,line:pos1}.
If an IHDR is generated with \texttt{color\_type} equal to \texttt{GrayScale}, the next preferred chunk will be IDAT (\Cref{line:pref2}), with several possible chunks available (\Cref{line:pos2}), including tIME and bKGD.
After IDAT is generated, IEND will be the next preferred chunk (\Cref{line:pref3}) and the \texttt{possible} array is updated with operators \texttt{-=} and \texttt{+=} in \Cref{line:pos3,line:pos32} to remove and add new values.
For instance, bKGD is removed because a bKGD chunk is not allowed after IDAT.
Finally, when an IEND is generated, both arrays become empty (\Cref{line:pref4,line:pos4}) to signal that \texttt{ReadBytes()} must return \texttt{false}.

Our implementation of \texttt{ReadBytes()} allows the same \texttt{while} loop from \Cref{lst:ReadBytes} to work for both parsing and generation.
In parsing mode, we can easily check whether the bytes read from the file belong to the \texttt{preferred} array, the \texttt{possible} array or to neither (for example, if the position of those bytes would be beyond the end of the file).
This allows us to successfully reconstruct the random decisions that would need to be made to generate such a file.
Our approach of using \texttt{ReadBytes()} to define chunk ordering turned out to be extremely general, being used in five of the developed formats (PNG, JPEG, MP4, ZIP, AVI).

\subsection{Compressed and Encoded Data}

With all the changes shown so far, our modified binary template is able to generate almost all chunks of a PNG file correctly.
Only the IDAT chunk is still a challenge, because it contains a zlib-compressed datastream of the image pixel data.
In order to handle compressed (or otherwise encoded) data, the compression and decompression functions should be specified such that the generator can generate the uncompressed data first and then apply the compression function to it to obtain the datastream that will be written to the file.
We have manually implemented this strategy for the PNG IDAT chunk by calling the appropriate functions from the zlib library, allowing \tool{} to generate fully valid PNGs.
PNG was the only format so far to require this special handling for compression, because interestingly we found that in other formats simply generating random bytes for the compressed datastreams was sufficient to obtain streams that could be successfully decompressed with high probability.

\section{Implementation}
\label{sec:compiler}

In the domain of language-specific fuzzers, \tool{} is a \emph{generator compiler,} bringing together the versatility of language specifications and the efficiency of compiled generator code. \tool{} \emph{compiles} a binary template to C++ code for generating and parsing inputs in the format specified by the template, leveraging existing libraries for parsing~\cite{py010parser} and processing~\cite{pfp} binary templates.
The generated C++ code for a given format can be compiled into a standalone executable or a shared library that can be loaded by other fuzzers, such as AFL++~\cite{AFLplusplus-Woot20}.
Each \texttt{struct} definition in the binary template is implemented as a C++ class.
We also create C++ classes to handle native types, such as \texttt{uint32} and arrays.
Every time a variable is declared in the template, this triggers a call to the \texttt{generate()} method of the corresponding class.
\tool{} supports all important features of binary templates, including indexing into previous instances of a given variable, and recursive structs, where one \texttt{struct} can contain another \texttt{struct} of the same type.
We support integer variables in big endian and little endian modes, as well as bitfield variables, which can occupy only fractions of a byte.

Our implementation allows setting a maximum size in bytes for the generated file.
This is important for fuzzing because smaller files are both faster to produce and to process with the target program.
In our experiments, we have set the maximum file size to 64 kB, since we found that even with only a few kilobytes the files can already exhibit the whole variety of chunk types and structures inside them.
During generation, any attempt to generate too much data, such as defining an array with a huge length, will throw an exception, aborting the generation.
Generation is also aborted if the generator attempts to consume more bytes than available in the decision seed or performs some invalid operation, such as accessing a non-existing field from a \texttt{struct} (this can happen for some formats, since control-flow decisions can determine which fields are generated inside a \texttt{struct}).

We have also integrated into \tool{} quality assurance tests to verify the correct behavior of our generators and parsers.
We use round-trip tests to make sure that whenever a file is successfully generated, then it can also be correctly parsed, and vice versa.
We also ensure that our fuzzers can correctly parse a corpus of real files from each format, which were manually collected from GitHub.

\section{Decision Seeds}
\label{sec:decisions}

The methodology followed by \tool{} is the following: To generate a file, we first provide an array of random bytes, called the \emph{decision seed}. This array is the source of randomness for generator decisions. Such decision seeds can also be reconstructed by \tool{} while \emph{parsing}.

\subsection{Using Decision Seeds for Generation}

\begin{figure}
\center
\includegraphics[width=10cm]{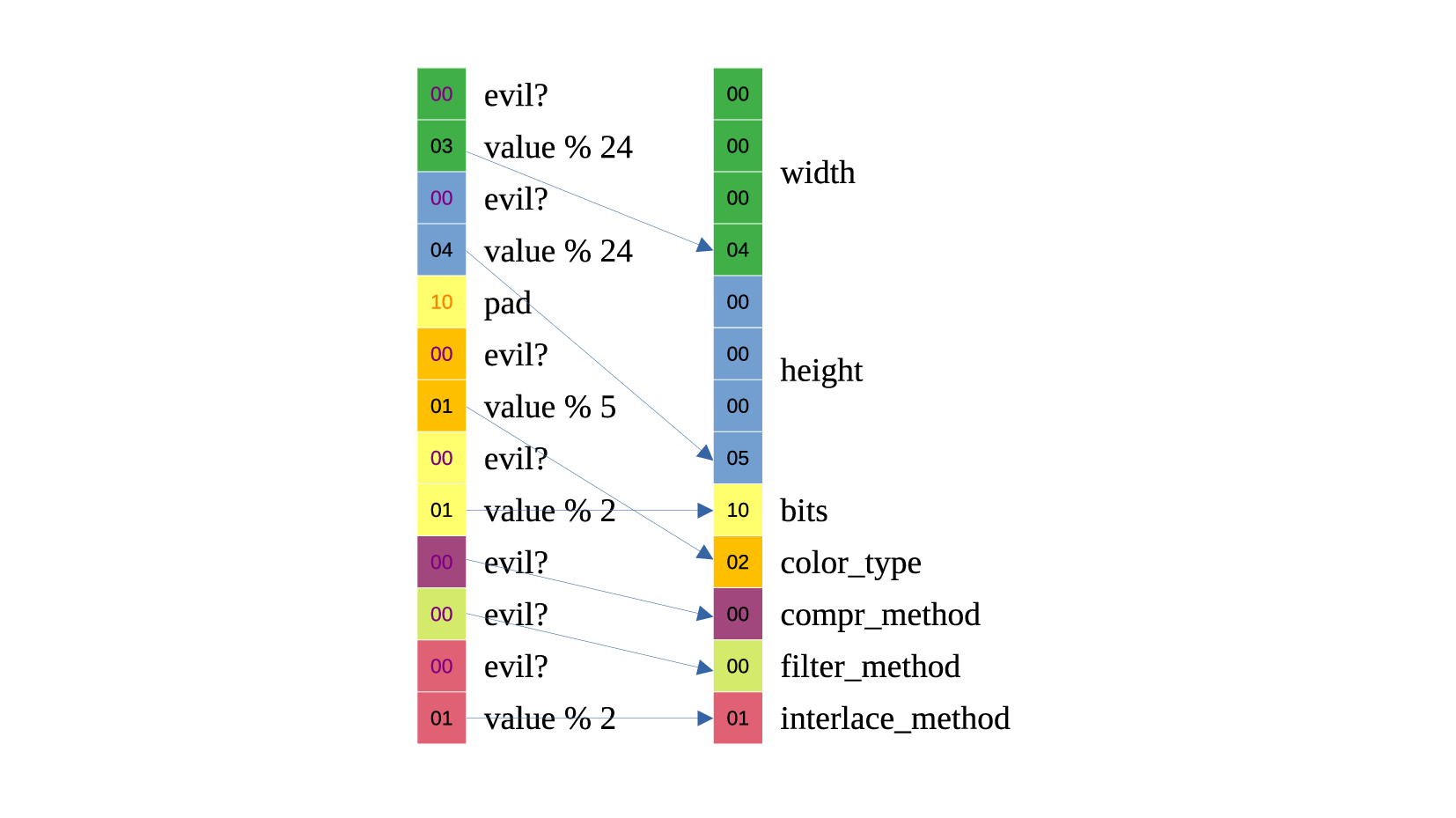}
\caption{Decision seed and the corresponding generated content for \texttt{PNG\_CHUNK\_IHDR} (defined in \Cref{lst:ihdr}).}
\label{fig:decisions}
\end{figure}

Every time the generator needs to make a random decision, such as which value should be used for a variable or lookahead function, it will read the required number of bytes from the decision seed.
For example, \Cref{fig:decisions} shows one possible decision seed and the resulting contents of the generated file for a \texttt{PNG\_CHUNK\_IHDR} (defined in \Cref{lst:ihdr}).
For each of the seven generated variables in this chunk, one decision byte is used to choose whether to make an evil decision.
If byte $b_0$ is read, an evil decision will be taken if $b_0\bmod 128 = 127$, which happens with probability $1/128$.
In the most common case, no evil decision is made.
For variables \texttt{width} and \texttt{height}, which were declared with \texttt{<min=1, max=24>}, this means that the next byte $b_1$ will be used to compute $b_1\bmod 24$ in order to choose uniformly from one of the 24 possible values.

For the \texttt{bits} and \texttt{color\_type} variables, the \texttt{ReadByte()} call in \Cref{line:ReadByte} of \Cref{lst:ihdr} tells us to first skip the \texttt{bits} variable and pick a suitable value for \texttt{color\_type} from the array \texttt{colors}, which are the valid values for a \texttt{PNG\_COLOR\_SPACE\_TYPE}.
In order to do that, we first consume one decision byte that will be used as a temporary padding for the byte we skipped in the file when jumping to \texttt{FTell()+1}.
Then, the next two decision bytes are used to pick the value at position \texttt{FTell()+1} (which will be \texttt{color\_type}): one to decide whether to make an evil decision and one to pick an appropriate value from the five possible choices in the array \texttt{colors}.
If we pick the second value in the array, which is \texttt{TrueColor} with value 2, then \Cref{line:truecolor} of \Cref{lst:ihdr} tells us that there are two possible values to pick for the \texttt{bits} variable: 8 or 16.
Finally, when we reach \Cref{line:color_type}, which defines the \texttt{color\_type} variable, we do not need to consume any more decision bytes, because its value had already been chosen at the previous call to \texttt{ReadByte()}.

Given a decision seed, the generation process is completely deterministic.
The decision seed encodes the essential features of a given file in a representation which is tailored for fuzzing by exhaustively exploring the file structure.
The correspondence between the decision seeds and the generated files can be thought of as a partial function $f:{\rm\it Seeds}\nrightarrow {\rm\it Files}$.
For some values of the seed, the generation can fail.
For example, say we have taken an evil decision such as an invalid \texttt{color\_type} in the PNG IHDR chunk.  This makes it impossible to later generate the data for another chunk.
But if the generation succeeds, it associates the seed $s$ to a unique well-defined file $f(s)$.
The resulting file $f(s)$ is not guaranteed to conform to the format specification (for instance, due to evil decisions), but it should be valid with high probability.
This mapping $f$ is not injective, but ideally every valid file should be obtainable by the generator, i.e. ${\rm\it Valid}\subseteq f({\rm\it Seeds})$.
This completeness property can be experimentally checked by attempting to parse valid files -- say, downloaded from the Internet -- with \tool{}.
As discussed in the next section, whenever parsing succeeds, we also obtain a decision seed for the parsed file, which in turn means that this file can be the output of the \tool{} generator by taking the appropriate decisions.

\subsection{Creating Decision Seeds During Parsing}
\label{sec:parsing}

When \tool{} runs in parsing mode, we obtain not only the parse tree for the input file, but also a \emph{decision seed} that can be used to generate the same input file, as presented in the diagram of \Cref{fig:overview}.
This is done by reconstructing at each step the decision bytes that would be required to produce the target file contents.
For example, in \Cref{fig:decisions}, the decision bytes on the left are exactly the ones that would be obtained by parsing the file contents on the right.
For each variable, we find out if an evil decision is required by checking whether its value is one of the possible values specified.
In the most common case, no evil decision is required and we can reconstruct the next decision byte as the index of the value that should be picked from the array of possible values.
This reconstructed decision seed is not unique.
For example, there are 24 valid values for the variables \texttt{width} and \texttt{height}, so any decision byte which is in the same congruence class modulo 24 would result in the same value.

One important axiom of our framework is that a file can be successfully parsed by \tool{} if and only if the file can be generated from some appropriate decision seed.
And the generator and parser must always agree on the same parse tree for the file.
That is why we cannot allow evil decisions when fixing the \texttt{length} field in \Cref{line:correct} of \Cref{lst:chunk}.
If the generator were to overwrite \texttt{length} with an evil value, the generator and the parser would disagree about the starting position of the next chunk.
Generally, the variables which define sizes or positions in the file, as discussed in \Cref{sec:size}, are the only ones which must be generated strictly without evil decisions.
Our round-trip tests have shown that generated files can be correctly parsed (and vice versa) even when all other variables are allowed to have evil values.

\section{Fuzzing Strategies}
\label{sec:strategies}

The \tool{} generators and parsers can be used for numerous fuzzing strategies. We discuss a few in this section.

\subsection{Generating Random Files}

The simplest approach to fuzzing with \tool{} is to use completely random seeds (say, from \texttt{/dev/urandom}) to generate files in a black-box manner.
A good fraction of the generated files will be valid according to the format specification and the files will come from a \emph{semantically diverse distribution} in terms of covered features.
For example, as discussed in \Cref{lst:ihdr}, all possible values for \texttt{color\_type} will be chosen with equal probability, as will the possible values of \texttt{bits} for a given \texttt{color\_type}.

\subsection{Mutating Inputs}
\label{sec:mutating}

\begin{figure}
\center
\includegraphics[width=6cm]{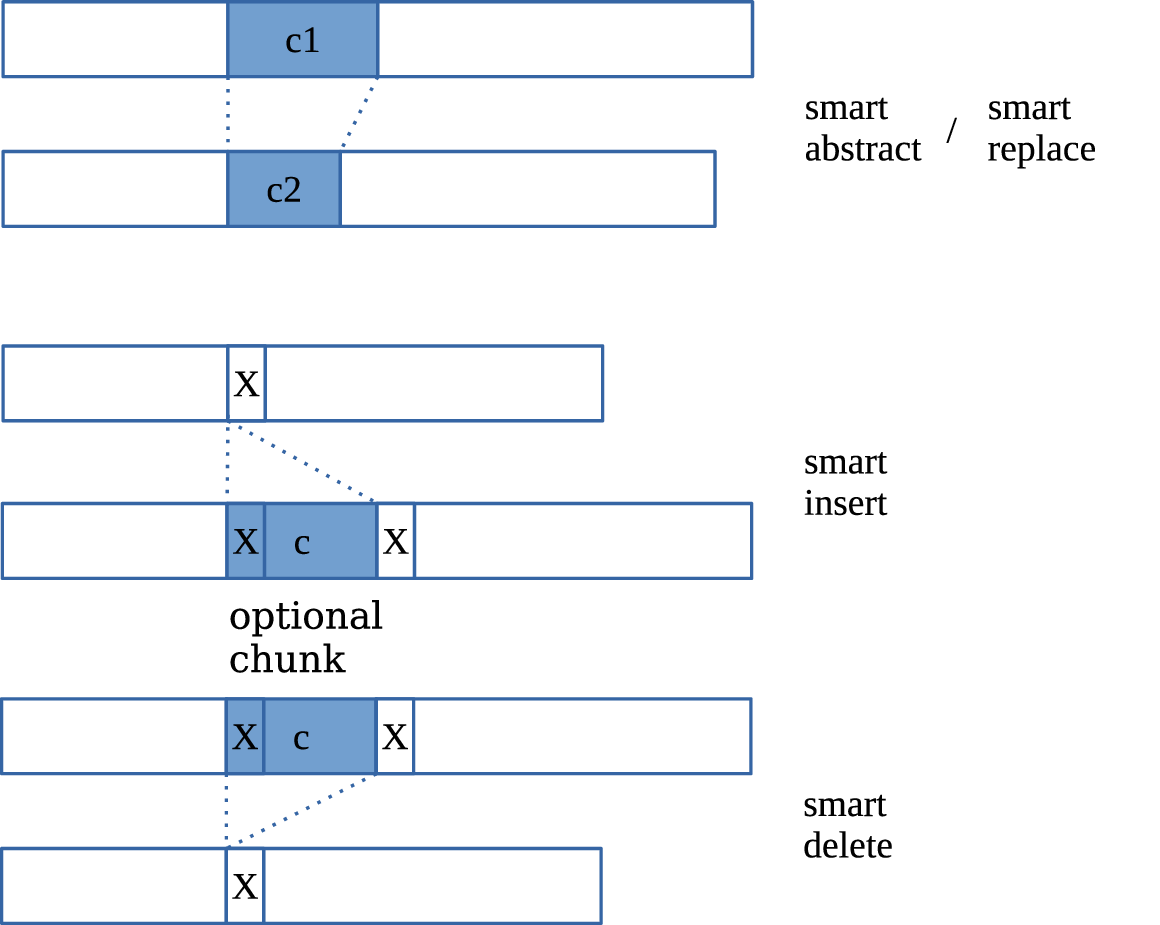}
\caption{Smart mutations in \tool{}. Here, X indicates the use of lookahead functions.}
\label{fig:smart}
\end{figure}

Another fuzzing strategy enabled by \tool{} is the use of \emph{smart mutations,} where chunks can be abstracted, replaced, deleted or inserted into a file, as seen in \Cref{fig:smart}.
Here, we use `chunk' generically to refer to any \texttt{struct} or variable defined in the binary template.
We define novel smart mutation operations which work over decision seeds, thus allowing contextual information to be taken into account when producing the mutated file.

\subsubsection{Smart Abstraction}
The first operator we discuss is \emph{smart abstraction}, which allows abstracting one specific chunk $c_1$ into a new random version $c_2$.
This operation works by first parsing the original file and obtaining its decision seed.
Then, we generate a new file by repeating all the decisions that were made before the target chunk $c_1$ and then using random bytes from \texttt{/dev/urandom} to produce the target chunk.
Once we detect that the target chunk has been generated, we go back to using the same decision bytes that were used after chunk $c_1$ in the original file.
We have found that smart abstractions were the most useful in reaching new coverage, and they are only possible because of our \tool{} framework, which enables synchronized generation and parsing.

\subsubsection{Smart Replacement}
The next operator \emph{smart replacement}, which replaces a chunk $c_1$ from ${\rm\it file}_1$ with a different chunk $c_2$ of the same type from ${\rm\it file}_2$.
Again, we first parse both files to obtain the decision seeds ${\rm\it seed}_1$ and ${\rm\it seed}_2$.
Then, a new mutated seed is created by replacing the decision bytes which created chunk $c_1$ with the decision bytes that create chunk $c_2$, leaving the remaining parts of the seed unchanged, as depicted in the \emph{smart replace} operation of \Cref{fig:smart}.
This mutated seed is then used by the generator to produce the mutated file.

To see why performing these operations on the decision seeds can help to generate valid files, consider the case where $c_1$ is some internal field of a PNG chunk.
If $c_1$ is replaced by a new instance~$c_2$ of the same type, but a different value and different size, then the generator will still compute the correct checksum and size for the modified chunk, while a simple copying of the contents of $c_2$ into $c_1$ would leave the checksum and size invalid.
This is especially important for successful smart mutations of formats such as MP4, which consist of structs called \emph{boxes} which recursively contain other \emph{boxes}.

\subsubsection{Smart Deletions}

Our \emph{smart delete} operation consists of deleting a chunk from a file, by removing the decision bytes which were responsible for the generation of this chunk.
Such a deletion operation only makes sense if we identify that the target chunk $c$ is optional.
We define a chunk $c$ to be \emph{optional} when, right before this chunk is generated, the template calls a lookahead function.
This signals that, depending on the result of this lookahead call, we might decide not to generate $c$ at all.
For example, the variable \texttt{chunk} declared in \Cref{line:chunk} of \Cref{lst:ReadBytes} is considered optional, because its declaration comes after a call to \texttt{ReadBytes()}.
We allow smart deletion to be performed over a chunk $c$ only when a lookahead function has been called right before and right after the generation of $c$, as shown in \Cref{fig:smart}.
This way, with the new mutated seed, the call to the lookahead function will consume the bytes that were originally consumed at the lookahead call that came after chunk $c$.

\subsubsection{Smart Insertions}

Conversely, we define a \emph{smart insert} as the inverse of the \emph{smart delete} operation, trying to insert a new chunk $c$ into a file.
Here, the original file must have a call to a lookahead function at the insertion position, so that the result of this call could be used to decide whether chunk $c$ should be created.
The inserted chunk $c$ must also be optional.
When performing smart insertions and replacements, \tool{} also checks whether the correct number of decision bytes has been  consumed in the generation of the new chunk.
In case the mutations do not work well because the new chunk does not fit in the desired position, this can be identified and reported by \tool{}.
For example, attempting to copy a bKGD chunk (which specifies a background color) will not work if one file uses color type \texttt{TrueColor}, which requires three values, while the other uses \texttt{GrayScale}, which requires only one.

\subsubsection{Cross-File Operations}

\tool{} implements procedures to parse a list of files and remember all the information about the chunks so they can be used later for smart mutations.
There is also a procedure to apply one smart mutation to a chosen file.
Here, we randomly choose which kind of mutation will be applied and which chunks will be involved.

\subsection{Integrating with Format-Agnostic Fuzzers}
\label{sec:integration}

Besides generation and mutation, \tool{} can also integrate with existing \emph{format-agnostic} fuzzers such as AFL~\cite{afl} in different ways.
For our experiments, we have integrated \tool{} with AFL++~\cite{AFLplusplus-Woot20} version 2.60c as follows.

\begin{description}
\item[\textsc{AFL+FFGen}] uses AFL to mutate and evolve the \emph{decision seeds}, which will be later fed into the \tool{} generator to produce inputs for the target program.
AFL can use the coverage feedback from the program to learn how to effectively mutate the decision seeds.
The advantage of working on such seeds is that they are simpler than binary files, since each byte corresponds to a unique decision, and those decisions are made sequentially in the order in which they appear in the seed.
Since the \tool{} generator already takes care of the correct input structure, such as computing the correct checksums, inserting the appropriate magic values and setting the correct size fields, AFL can focus exclusively on the high-level decisions that represent the file.
\item[\textsc{AFL+FFMut}] lets AFL mutate the files that will be given to the target program as usual, but also adds new mutation operations which are the smart mutations provided by \tool{}.
Every interesting input, which achieves new coverage, is saved in the AFL queue.
When such an input is about to be mutated for fuzzing, we also parse this input with \tool{}, allowing it be used for smart mutations.
\end{description}

\section{Evaluation}
\label{sec:evaluation}

Our evaluation focuses on the following research questions:

\def\rqone{How much effort is it to set up a binary template file for input generation?}

\def\rqtwo{How \emph{efficient} is \tool{} in producing inputs?}

\def\rqthree{How \emph{accurate} is \tool{} in producing inputs?}

\def\rqfour{How accurate are the smart mutations applied by \tool{}?}

\def\rqfive{How efficient is \tool{} as a standalone black-box fuzzer?}

\def\rqsix{How efficient is the integration of \tool{} with a format-agnostic fuzzer?}

\def\rqseven{How does \tool{} compare against other format-aware fuzzers?}

\def\rqeight{Does \tool{} find real bugs?}

\begin{description}
\item[RQ1] \rqone
\item[RQ2] \rqtwo
\item[RQ3] \rqthree
\item[RQ4] \rqfour
\item[RQ5] \rqfive
\item[RQ6] \rqsix
\item[RQ7] \rqseven
\item[RQ8] \rqeight
\end{description}

To answer these questions, we ran a series of experiments. All our experiments were conducted on an Intel Xeon CPU E5-4650L machine with 64 cores and 756 GB RAM, running Debian 10.
The fuzzers were run on a single processor for 24 hours.
We have repeated each experiment for a total of 10 runs.

In our evaluation, we compare \tool{} against the format-agnostic fuzzer AFL++~\cite{AFLplusplus-Woot20}, as well as the state-of-the-art fuzzer of binary file formats AFLSmart~\cite{pham2019smart}.
We do not compare against the grammar-based fuzzers Superion~\cite{wang2019superion}, Nautilus~\cite{aschermann2019nautilus} and Grimoire~\cite{blazytko2019grimoire} because such fuzzers were only evaluated on text-based grammar input formats, such as markup languages (XML) or programming languages (JavaScript, PHP, Ruby, Lua, C, nasm, SQL, SMT).
These textual formats do not have common features of binary formats, such as size fields, checksums, or bitfields.
Therefore, it is not clear how well such fuzzers could handle those features.
For Superion and Nautilus, a complete format specification would have to be written from scratch for each new binary format.
On the other hand, Grimoire can mine format specifications, but it only supports textual languages, as it splits its inputs based on particular characters, such as opening and closing brackets and quotation marks.
The Grimoire paper points out, for instance, that the tool can apply effective mutations to Lua source code, but not to Lua bytecode.

For evaluations, we have chosen 10 popular binary file formats, including the most popular formats for compressed archives (ZIP), images (PNG, JPG) and videos (MP4), as well as additional formats which are supported by AFLSmart.

\def\colhead{\color{white}}

\begin{table}  %
  \rowcolors{2}{rltblue!10}{white}
  \centering
  \caption{Number of lines of code required for each format.}
  \small
  \label{tab:lines}
  \begin{tabular}{lrrrlr}
  \rowcolor{rltblue}
  \colhead Format & \colhead Original & \colhead Modified Template & \multicolumn{2}{c}{\colhead Changes} & \colhead Generated C++ \\
    GIF & 204 & 234 & $+38$ & $-8$ & 2,144  \\
BMP & 138 & 249 & $+130$ & $-19$ & 1,267  \\
PCAP & 220 & 268 & $+72$ & $-24$ & 1,388  \\
MIDI & 261 & 282 & $+39$ & $-18$ & 1,771  \\
AVI & 304 & 471 & $+248$ & $-81$ & 2,645  \\
PNG & 388 & 492 & $+114$ & $-10$ & 2,664 (+170)  \\
WAV & 574 & 649 & $+108$ & $-33$ & 2,457  \\
ZIP & 669 & 774 & $+191$ & $-86$ & 2,608  \\
MP4 & 807 & 1,420 & $+771$ & $-158$ & 4,556  \\
JPG & 1,631 & 1,662 & $+282$ & $-251$ & 5,683
  \end{tabular}
\end{table}

\subsection{RQ1: Effort for Extending Template Files}
\label{sec:loc}

We start with RQ1: \emph{\rqone} To answer this question, \Cref{tab:lines} lists the number of lines of code required for each format specification.
We list the sizes of the original binary template and modified version, which supports generation.
We also detail the number of lines which were added or removed.
For most formats the number of changes is small compared to the amount of code we can already leverage from existing parsing-only binary templates.

All of the formats shown in \Cref{tab:lines}, with the exception of PNG, were developed by three different BSc students, who did not have prior experience with binary formats.
Once they had learned the \emph{binary template language} while creating the first format, development was easier, since many patterns are common to multiple formats.
From their experience, a couple of days is sufficient for updating most formats for generation.
However, a few complex formats took more than one week to modify.
That was the case with MP4, which is composed of several different chunk types, many of which were not fully described in the original binary template.
Note, though, that this effort can hardly be avoided: A random fuzzer will only very occasionally generate a valid MP4 file, let alone systematically cover all chunk types; and a hand-coded MP4 generator will require the same or higher specification effort (and neither result in a parser nor a mutator).

\Cref{tab:lines} also lists the number of lines of format-specific C++ code which were automatically generated by \tool{}.
We have found that the use of a domain-specific language of binary templates makes our specifications much more succinct and readable than if they were developed from scratch in a general-purpose language such as C++.
Only for the PNG format did we have to edit the C++ code directly to support data compression; in the future, we plan to add native support for compression in \tool{}, making such manual changes unnecessary.

To give some perpective, the first author of this paper had written a PNG generator completely from scratch in C++, before seeing the PNG binary template~\cite{010editortemplatesrepository}, by reading the PNG~specification~\cite{png}, a document which spans 50 pages.
This manually written implementation had 596~lines of C++ code, which is a lot higher than the number of lines of code that had to be modified by starting from an existing PNG binary template and leveraging the \tool{} framework---and it is a generator only, without any parsing or mutation capabilities, let alone coverage guidance.

\takeaway{\label{take:effort}Extending binary templates is less work than writing a generator from scratch, and provides a parser and mutator on top.}

\begin{table}
  \rowcolors{2}{rltgreen!10}{white}
  \caption{Average performance of the fuzzers in terms of speed and validity.}
  \footnotesize
  \label{tab:performance}
  \begin{tabular}{lrrrrrrrrl}
    \rowcolor{rltgreen}
 & \multicolumn{2}{c}{\colhead Speed (files / $s$)} & \colhead Size & \multicolumn{2}{c}{\colhead With Evil} & \multicolumn{2}{c}{\colhead Without Evil} & \colhead AFL & \\
 \rowcolor{rltgreen}
 \colhead Fmt. & \colhead Gen. & \colhead Parsed & \colhead ($B$) & \colhead Suc. & \colhead Valid & \colhead Suc. & \colhead Valid & \colhead Valid & \colhead Validation Command \\
PNG & 6,229 & 2,488 & 752 & 97\% & 84\% & 100\% & 92\% & 5\% & \scriptsize\texttt{identify -verbose - | grep Elapsed}  \\
JPG & 7,457 & 5,059 & 1,343 & 100\% & 81\% & 100\% & 95\% & 66\% & \scriptsize\texttt{identify -verbose - | grep Elapsed}  \\
GIF & 6,111 & 3,760 & 6,007 & 100\% & 40\% & 100\% & 40\% & 14\% & \scriptsize\texttt{identify -verbose - | grep Elapsed}  \\
MIDI & 3,453 & 21,639 & 573 & 96\% & 74\% & 100\% & 74\% & 60\% & \scriptsize\texttt{! timidity - -Ol | grep }$\wedge$\texttt{-:}  \\
MP4 & 3,786 & 2,465 & 2,093 & 99\% & 82\% & 100\% & 88\% & 34\% & \tiny\texttt{ffmpeg -y -i - -c:v mpeg4 -c:a copy o.mp4}  \\
ZIP & 10,229 & 13,844 & 1,894 & 96\% & 72\% & 98\% & 75\% & 2\% & \scriptsize\texttt{yes | unzip -P "" -t }\textnormal{$\langle$input file$\rangle$}\footnotemark \\
PCAP & 6,736 & 5,299 & 2,190 & 95\% & 91\% & 100\% & 100\% & 3\% & \scriptsize\texttt{tcpdump -nr -}  \\
AVI & 18,489 & 25,560 & 415 & 95\% & 84\% & 100\% & 90\% & 27\% & \scriptsize\texttt{ffmpeg -y -f avi -i - output.avi} \\
WAV & 7,279 & 7,988 & 4,014 & 95\% & 96\% & 98\% & 99\% & 26\% & \scriptsize\texttt{wavpack -y - -o output.wav} \\
BMP & 4,377 & 6,225 & 1,173 & 95\% & 80\% & 95\% & 83\% & 23\% & \scriptsize\texttt{identify -verbose - | grep Elapsed} \\
  \end{tabular}
\end{table}

\footnotetext{For \texttt{unzip}, both exit status 0 (normal) and exit status 1 (just warnings, but no errors) are considered valid files.}

\subsection{RQ2: Generator Speed}
\label{sec:speed}

For RQ2 (\emph{\rqtwo}), we measure the speed of our generators and parsers and classify the validity of the generated files by feeding them to a target program.
The question of how to define the validity of files can be somewhat tricky.
Different programs that process a given format tend not to agree about which inputs they accept, and they often do not follow the official specification.
Programs such as media players tend to be forgiving and continue processing the input even after identifying some data corruption.
In line with our main goal of reaching deeper code in the application, we classify a file as invalid only when it triggers a critical error which prevents further processing of the file.
For example, for images we used the command \texttt{identify -verbose} from ImageMagick and checked the output to see if the program could successfully print detailed information about the image, even if some errors were reported.

\Cref{tab:performance} summarizes our results.
For each format-aware fuzzer, we have generated 10,000 files, then also parsed those files and tested them with a target program to check for validity (validation command).
Here, we see that \tool{} can both generate and parse inputs at a speed of thousands per second for all the tested file formats.
This is comparable or sometimes even faster than the speed with which format-agnostic fuzzers, such as AFL, can execute inputs.
Therefore, our generators and parsers can be easily integrated into existing fuzzers without becoming a performance bottleneck.

\takeaway{\tool{} is very fast, generating and parsing thousands of inputs per second.}

\subsection{RQ3: Input Validity}
\label{sec:validity}

\Cref{tab:performance} also shows which fraction of the generations produced an output (successfully completed generation) and which fraction of those outputs was classified as valid, addressing RQ3: \emph{\rqthree}
The results show that our fuzzers succeed almost all the time in producing inputs and those inputs have a high chance of being accepted by the target program.
The validity is even higher if evil decisions are disabled.
However, evil decisions are particularly useful in fuzzing to trigger some unexpected behaviors.

\takeaway{The large majority of files generated by \tool{} is valid.}

To put this validity into perspective, let us compare against inputs produced by the format-agnostic AFL fuzzer, listed under the ``AFL'' column.
We see that even with sample files, AFL produces far fewer valid input files.
In fact, if we compute the ratio of valid files produced by \tool{} and valid files produced by AFL, we obtain geometric average of 4.9 (minimum 1.2, maximum 36.0).

\takeaway{\tool{} produces on average five times more valid inputs than a format-agnostic fuzzer.}

In the next section, we also compare the validity of inputs produced by \tool{} against inputs produced by AFLSmart.

\subsection{RQ4: Mutation Validity}
\label{sec:accuracy}

For RQ4 (\emph{\rqfour}), we evaluate how successful our smart mutations are compared to simpler mutations which do not use decision seeds.
From our four smart mutations defined in \Cref{sec:mutating}, three of them (\emph{replace}, \emph{insert}, \emph{delete}) could be implemented in a \emph{simple} way by applying the corresponding operation (\emph{replace}, \emph{insert}, \emph{delete}) directly to the chunks of the files, without using decision seeds.
Such implementation is common in prior work, such as in AFLSmart~\cite{pham2019smart}.

In order to evaluate our smart mutations, we have applied each type of mutation 10,000 times over the initial corpus of valid files for each format.
\Cref{tab:mutations} show our results, where the numbers are expressed as percentages.
For each smart mutation, we first show the success rate (which percentage of tries successfully generated a file).
For the \emph{replace} and \emph{insert} mutations, we also show in parenthesis the percentage of tries for which we consumed exactly the expected number of decision bytes while generating the target chunk.
In those cases, we should expect the resulting file to be the semantically correct result of applying the desired mutation.
But even when this does not happen (for example, if the chunk to be inserted cannot fit in the target position), in the great majority of cases we still manage to produce a file, which is helpful for fuzzing.
In fact, the success rate in generating a file is almost always above 91\%.

From those cases where a file is successfully generated, we report which percentage of those files are considered valid, and also which percentage would be valid if we applied the \emph{simple} version of the corresponding mutation.
As shown in \Cref{tab:mutations}, we obtain almost universally higher validity when applying our smart mutations over decision seeds, when compared to their \emph{simple} versions.
This is expected, as our smart mutations are able to take contextual information into account, for example, being able to recompute the correct checksum for a chunk, even when the contents of the chunk have been partially modified.
We also report on which percentage of the cases the files resulting from our smart mutations are different from what would be obtained by applying a \emph{simple} mutation.
Finally, we would also like to point out that our \emph{smart abstract} mutations are only possible due to \tool{}'s ability to generate new chunks completely from scratch, and those abstractions produce very high validity, being the most profitable type of mutation in reaching new coverage during fuzzing.

\begin{table}
  \rowcolors{2}{rltgreen!10}{white}
  \caption{Smart mutations: success (Suc.), semantic correctness of the target chunk (Cor.), validity (Val.), validity of the corresponding \emph{simple} mutation (Sim.), and fraction of cases where our smart mutations differ from the \emph{simple} versions (Diff.).  All numbers are expressed as percentages (\%).}
  \footnotesize
  \label{tab:mutations}
  \begin{tabular}{lrrrrrrrrrrrrrr}
    \rowcolor{rltgreen}
 & \multicolumn{4}{c}{\colhead Replace} & \multicolumn{4}{c}{\colhead Insert} & \multicolumn{4}{c}{\colhead Delete} & \multicolumn{2}{c}{\colhead Abstract} \\
 \rowcolor{rltgreen}
 \colhead Fmt. & \colhead Suc. (Cor.) & \colhead Val. & \colhead Sim. & \colhead Diff. & \colhead Suc. (Cor.) & \colhead Val. & \colhead Sim. & \colhead Diff. & \colhead Suc. & \colhead Val. & \colhead Sim. & \colhead Diff. & \colhead Suc. & \colhead Val.  \\
PNG & 98 (36) & 84 & 7 & 98 & 95 (12) & 60 & 19 & 97 & 96 & 75 & 37 & 73 & 100 & 90  \\
JPG & 100 (58) & 84 & 68 & 70 & 100 (38) & 45 & 79 & 87 & 100 & 86 & 86 & 50 & 100 & 83  \\
GIF & 100 (61) & 39 & 29 & 51 & 100 (60) & 24 & 20 & 49 & 100 & 6 & 6 & 9 & 100 & 41  \\
MIDI & 100 (91) & 99 & 92 & 45 & 100 (98) & 100 & 99 & 100 & --/--\footnotemark & --/-- & --/-- & --/-- & 100 & 98  \\
MP4 & 94 (39) & 67 & 43 & 77 & 86 (16) & 24 & 19 & 98 & 77 & 30 & 28 & 99 & 96 & 71  \\
ZIP & 99 (54) & 77 & 27 & 81 & 96 (41) & 71 & 29 & 100 & 99 & 77 & 0 & 100 & 99 & 83  \\
PCAP & 99 (86) & 100 & 88 & 24 & 100 (100) & 100 & 100 & 0 & 100 & 100 & 100 & 0 & 98 & 99  \\
AVI & 91 (53) & 78 & 75 & 80 & 92 (37) & 77 & 86 & 100 & 91 & 89 & 84 & 100 & 94 & 86  \\
WAV & 97 (45) & 74 & 42 & 85 & 92 (27) & 75 & 33 & 99 & 95 & 61 & 31 & 97 & 97 & 78  \\
BMP & 97 (58) & 95 & 61 & 84 & --/--\footnotemark & --/-- & --/-- & --/-- & --/-- & --/-- & --/-- & --/-- & 99 & 93  \\
  \end{tabular}
\end{table}
\footnotetext[2]{The MIDI format does not have any deletable chunks.}
\footnotetext{The BMP format does not have any optional chunks, so insertions and deletions are not possible.}

\takeaway{Smart mutations applied over decision seeds, as performed by \tool{}, are far more likely to yield valid inputs than smart mutations which simply copy the contents of chunks from the files, as the ones used by prior work, such as AFLSmart.}

\subsection{RQ5: Black-Box Fuzzing}
\label{sec:blackbox}

Let us now evaluate the \emph{effectiveness} of \tool{}, notably in achieving \emph{coverage} in our test subjects. Generally speaking, the higher the coverage of a test generator, the higher its chance to detect bugs---in particular, since if code is not covered, bugs will not be detected.

We start with RQ5: \emph{\rqfive} We evaluate two black-box settings:
\begin{description}
\item[\textsc{FFGen}] is using \tool{} as a standalone input generator, requiring no knowledge or feedback from the program under test. (Note that this is a setting in which only a specification-based fuzzer like \tool{} can succeed.) 
\item[\textsc{FFMut}] is using \tool{} to parse one initial set of input files and apply smart mutations, again without guidance.
\end{description}

\def\evalhead{
\colhead Setting & \colhead PNG & \colhead JPG & \colhead GIF & \colhead MIDI & \colhead MP4 & \colhead ZIP & \colhead PCAP & \colhead AVI & \colhead WAV & \colhead BMP}

First, we evaluate the \emph{language coverage}, i.e., which features of the binary template are actually present in the generated files.
In order to do this, we have generated 10,000 files with the black-box generation strategy \textsc{FFGen} and measured which percentage of variable declaration statements in the binary template have been covered, since variable declarations are the places where a new node is added to the parse tree.
\Cref{tab:language} shows the resulting language coverage, which is at least 94\% for almost all formats.
Only the JPG format had a lower coverage of 79\%.
The missing coverage was mostly due to some optional chunks which were enabled for parsing, but had not yet been modified to generate valid contents.
We generate such chunks only with very low probability, in order to ensure most generations will produce valid files.

\takeaway{\label{take:language}By itself, \tool{} extensively covers the language features present in \emph{valid} inputs.}

\begin{table}
  \rowcolors{2}{rltblue!10}{white}
  \centering
  \caption{Programs under test.}
  \small
  \label{tab:targets}
  \begin{tabular}{lll}
  \rowcolor{rltblue}
  \colhead Format & \colhead Program & \colhead Invocation \\
PNG & \texttt{libpng 1.6.37-3} & \texttt{readpng}  \\
JPG & \texttt{libjpeg-turbo 1\_2.0.6-4} & \texttt{djpeg \textnormal{$\langle$input file$\rangle$}} \\
GIF & \texttt{gif2png 2.5.14} & \texttt{gif2png \textnormal{$\langle$input file$\rangle$}} \\
MIDI & \texttt{TiMidity++ 2.14.0-8} & \texttt{timidity - -o - -Ow}  \\
MP4 & \texttt{FFmpeg 4.4} & \texttt{ffmpeg -y -i - out.mp4}  \\
ZIP & \texttt{UnZip 6.0-26} & \texttt{unzip -P "" -t \textnormal{$\langle$input file$\rangle$}} \\
PCAP & \texttt{tcpdump 4.99.0-2}, & \texttt{tcpdump -nr -}  \\
\rowcolor{white}
& \texttt{libpcap 1.10.0-2} & \\
\rowcolor{rltblue!10}
AVI & \texttt{FFmpeg 4.4} & \texttt{ffmpeg -y -i - out.avi} \\
\rowcolor{white}
WAV & \texttt{WavPack 5.4.0-1} & \texttt{wavpack -y - -o -} \\
\rowcolor{rltblue!10}
BMP & \texttt{libgdk-pixbuf 2.42.6-1} & \texttt{gdk-pixbuf-thumbnailer \textnormal{$\langle$input file$\rangle$} out.png} \\
\end{tabular}
\end{table}

\begin{table}
\rowcolors{2}{white}{rltgreen!10}
\caption{Language Coverage: coverage of variable declarations (\%) in the binary template.}
\label{tab:language}
\small
\begin{tabular}{@{}lrrrrrrrrrr@{}}
\rowcolor{rltgreen}
\evalhead \\
\textsc{FFGen} & 100 & 79 & 100 & 100 & 94 & 94 & 100 & 98 & 97 & 98 \\
\end{tabular}
\end{table}

Next, we evaluate how well the fuzzer performs on real-world applications.
\Cref{tab:targets} lists the programs with the exact command used for fuzzing.
The fuzzing experiments used a timeout of 24 hours.
For all techniques which require an initial corpus of inputs, we have used the same corpus of files, which were small files from each format manually downloaded from GitHub.
Each corpus had an average of 11 files (minimum 5, maximum 31).

\begin{table}[H]
\rowcolors{2}{brown!10}{white}
\caption{Line coverage (\%) of \tool{} in black-box settings. Here, $A\setminus B$ represents the lines which were covered by technique $A$, but not $B$.}
\label{tab:blackbox-gen}
\small
\begin{tabular}{@{}lrrrrrrrrrr@{}}
\rowcolor{brown}
\evalhead \\
\textsc{FFGen} & 23.0 & 24.6 & 68.9 & 12.4 & 5.7 & 34.8 & 12.5 & 5.8 & 21.8 & 27.8 \\
\textsc{FFMut} & 24.7 & 24.4 & 70.5 & 10.4 & 7.4 & 36.1 & 9.8 & 7.1 & 21.9 & 27.8 \\ \hline
$\textsc{FFGen} \setminus \textsc{FFMut}$ & 0.2 & 0.2 & 0.5 & 2.1 & 0.5 & 0.2 & 4.2 & 0.8 & 0.1 & 0.1  \\
$\textsc{FFMut} \setminus \textsc{FFGen}$ & 1.9 & 0.1 & 2.2 & 0.1 & 2.2 & 1.5 & 1.5 & 2.0 & 0.2 & 0.0  \\
\end{tabular}
\end{table}

\Cref{tab:blackbox-gen} lists the resulting line coverage obtained with LCOV (average over all runs) for the black-box settings. One interesting aspect is that in both settings, all inputs are generated syntactically valid by construction; and indeed, inspection shows that error-handling code is hardly covered.\footnote{Generally speaking, reaching error-handling code is the easy part of fuzzing; if one wants to cover error-handling code, too, one could easily introduce subtle lexical mutations into the files generated by \tool{}.}

\takeaway{\tool{} achieves decent coverage even in black-box settings, without any guidance from the program under test.}

The lower half of \Cref{tab:blackbox-gen} highlights the \emph{coverage difference} between the two settings; here, $A \setminus B$ denotes lines covered in~$A$, but not in $B$. The advantage of \textsc{FFGen} over \textsc{FFMut} is prominent in MIDI and PCAP, where the binary template covers exotic contents not found in our sample files.
Conversely, the additional coverage of \textsc{FFMut} over \textsc{FFGen} is likely due to some features in the corpus of input files which are not described in the binary templates.

\takeaway{Some features in the spec are unlikely to be found in files in the wild.}

Additionally, we have noticed that for all formats, the final coverage obtained after generating files with the \textsc{FFGen} approach exceeds the coverage from our downloaded corpus (by a factor of $1.4\times$ on average).

\subsection{RQ6: Integrating Format-Agnostic Fuzzers}
\label{sec:afl}

In practice, we will frequently encounter settings where feedback from the program under test is available, and where \tool{} therefore can integrate with existing format-agnostic fuzzers. Let us thus address RQ6: \emph{\rqsix}
We have integrated \tool{} with the popular AFL++~\cite{AFLplusplus-Woot20} fuzzer in two different ways, as presented in \Cref{sec:integration}:

\begin{description}
\item[\textsc{AFL+FFGen}] is using AFL to mutate the \emph{decision seed} which is then fed into \tool{} and used to generate inputs for the target program.
Since only the decision seed is mutated, the generated inputs should conform to the given format.

\item[\textsc{AFL+FFMut}] is having AFL use \tool{} to perform smart mutations on the input file.
\tool{} parses the files which are about to be fuzzed and remembers information about the chunks present in those files, which it can later use to perform smart mutations.
Note that all regular AFL mutations are also used, which can yield invalid inputs and helps covering error handling code.
The smart mutations from \tool{} are added as an additional custom mutator.
\end{description}

\begin{table}[h]
\rowcolors{2}{rltred!10}{white}
\caption{Line coverage (\%) of \tool{} integrated with AFL. Here, $A\setminus B$ represents the lines which were covered by technique $A$, but not $B$. Vargha-Delaney effect size $A^{12}$ (statistically significant effects in \textbf{bold}).}
\label{tab:afl}
\small
\begin{tabular}{lrrrrrrrrrr}
\rowcolor{rltred}
\evalhead \\
\textsc{AFL} & 20.7 & 29.8 & 73.3 & 13.3 & 11.5 & 36.9 & 25.3 & 11.2 & 22.4 & 30.7 \\
\textsc{AFL+FFGen} & 25.4 & 27.2 & 71.7 & 12.1 & 9.7 & 38.4 & 22.1 & 10.3 & 21.9 & 27.9 \\
\textsc{AFL+FFMut} & 27.8 & 33.6 & 73.3 & 14.3 & 11.4 & 38.6 & 26.2 & 11.8 & 22.3 & 30.7 \\
\hline
$\textsc{AFL+FFGen} \setminus \textsc{AFL}$ & 6.3 & 2.4 & 0.1 & 0.3 & 0.6 & 2.4 & 1.1 & 0.7 & 0.0 & 0.0  \\
$\textsc{AFL+FFMut} \setminus \textsc{AFL}$ & 7.2 & 3.8 & 0.0 & 1.7 & 0.9 & 2.1 & 2.3 & 1.0 & 0.1 & 0.0  \\
$\textsc{AFL} \setminus \textsc{AFL+FFGen}$ & 1.5 & 5.0 & 1.8 & 1.5 & 2.5 & 0.9 & 4.3 & 1.6 & 0.5 & 2.9  \\
$\textsc{AFL} \setminus \textsc{AFL+FFMut}$ & 0.0 & 0.0 & 0.1 & 0.7 & 0.9 & 0.4 & 1.5 & 0.5 & 0.3 & 0.0  \\
$\textsc{AFL+FFGen} \setminus \textsc{FFGen}$ & 2.5 & 2.8 & 3.3 & 0.2 & 4.1 & 3.6 & 10.0 & 4.4 & 0.3 & 0.0  \\
$\textsc{AFL+FFMut} \setminus \textsc{FFMut}$ & 3.2 & 9.2 & 2.9 & 3.9 & 4.1 & 2.6 & 16.5 & 4.5 & 0.4 & 2.9  \\
\hline
$A^{12}(\textsc{AFL+FFMut} > \textsc{AFL})$ & \textbf{1.0} & \textbf{1.0} & 0.34 & 0.66 & 0.52 & \textbf{1.0} & \textbf{0.81} & \textbf{0.87} & 0.41 & 0.45 \\
\end{tabular}
\end{table}

\Cref{tab:afl} lists the resulting coverage.
In our fuzzing experiments, we have enabled AFL dictionaries whenever suitable, for a fair comparison against the optimal setup of AFL.
We have not used dictionaries for the \textsc{AFL+FFGen} approach, since this fuzzer mutates decision seeds, and not binary files directly.
Dictionaries were enabled for all formats, with the exception of MIDI, for which there was no dictionary provided by AFL++.

For most subjects, the set differences show that both strategies increase coverage over plain AFL, showing that \tool{} is effective in making format-agnostic fuzzers format aware. For six out of ten subjects, the integration also achieves a higher coverage in absolute terms.

\Cref{tab:afl} also reports the Vargha-Delaney effect size $A^{12}$, which estimates the probability that \textsc{AFL+FFMut} will perform better than the competing technique \textsc{AFL}.
Values greater than 0.5 indicate that \textsc{AFL+FFMut} more likely to win this coverage comparison.
Statistically significant effects are marked in \textbf{bold}, according to the Wilcoxon signed-rank test.
As we can see, \textsc{AFL+FFMut} performed significantly better on five benchmarks, while for the remaining five there was no statistically significant difference.
On three benchmarks, \textsc{AFL} had a slight advantage.
The reason here was because the smart mutations from \tool{} require slightly more computation time than format-agnostic mutations, so \textsc{AFL} achieved more coverage simply by being able to run more executions.
But the advantage \textsc{AFL} had on those benchmarks is not statistically significant.

\takeaway{Format-aware fuzzing with \tool{} reaches lines that format-agnostic fuzzing does not.}

However, AFL by itself also covers lines that the integration of AFL and \tool{} did not reach. One reason for this is again AFL producing several \emph{invalid} inputs, covering error-handling code, which $\textsc{AFL+FFGen}$ avoids by construction.
The evil decisions in \tool{} do enable a limited class of invalid inputs, but such evil decisions are sparse and are not allowed to completely destroy the structure of the input, for example, by generating inputs that fail to parse due to an incorrect chunk size.
So the random mutations performed by AFL are still needed to cover all the parsing errors in the target program.

\takeaway{\label{take:complement}Format-aware and format-agnostic fuzzing complement each other.}

Comparing the AFL integration against \tool{} standalone, we also see that integrating AFL's coverage guidance into format-aware fuzzing also significantly improves coverage over a pure black-box setting. Hence, such integration is a preferred setting if sample inputs and coverage feedback are available.

\takeaway{Integrating \tool{} with coverage-guided fuzzers improves coverage over black-box settings.}

\subsection{RQ7: Alternate Format-Aware Strategies}
\label{sec:aflsmart}

Our next evaluation concerns RQ7: \emph{\rqseven} The competitor here is AFLSmart~\cite{pham2019smart}, which uses format information from \emph{Peach} specifications to determine chunk boundaries in input files, thus also resulting in smarter mutations. AFLSmart uses input specifications to \emph{parse} inputs only, but not to \emph{generate} valid inputs.

\begin{table}[h]
\rowcolors{2}{RawSienna!10}{white}
\caption{Line coverage (\%) of \tool{} and AFLSmart. Here, $A\setminus B$ represents the lines which were covered by technique $A$, but not $B$. Vargha-Delaney effect size $A^{12}$ (statistically significant effects in \textbf{bold}).}
\label{tab:aflsmart}
\small
\begin{tabular}{lrrrrrrrrrr}
\rowcolor{RawSienna}
\evalhead \\
\textsc{AFLSmart} & 20.7 & 29.7 & --/--\footnotemark & 14.1 & 12.3 & 36.4 & 26.4 & 12.2 & 22.2 & --/--\footnotemark \\
\textsc{AFL+FFMut} & 27.8 & 33.6 & 73.3 & 14.3 & 11.4 & 38.6 & 26.2 & 11.8 & 22.3 & 30.7 \\
\hline
$\textsc{AFLSmart} \setminus \textsc{AFL+FFMut}$ & 0.0 & 0.1 & --/-- & 1.7 & 1.3 & 0.2 & 1.9 & 1.0 & 0.0 & --/-- \\
$\textsc{AFL+FFMut} \setminus \textsc{AFLSmart}$ & 7.1 & 4.0 & --/-- & 1.9 & 0.5 & 2.4 & 1.7 & 0.6 & 0.1 & --/-- \\
\hline
$A^{12}(\textsc{AFL+FFMut} > \textsc{AFLSmart})$ & \textbf{1.0} & \textbf{1.0} & --/-- & 0.48 & \textbf{0.04} & \textbf{1.0} & 0.39 & \textbf{0.12} & \textbf{0.73} & --/-- \\
\end{tabular}
\end{table}
\footnotetext[5]{AFLSmart crashed with a segmentation fault.}
\footnotetext{BMP format is still not supported by AFLSmart.}

The coverage results are listed in \Cref{tab:aflsmart}. Comparing \textsc{AFLSmart} against \textsc{AFL+FFMut}, we see that the AFL+\tool{} integration outperforms \textsc{AFLSmart} in five out of eight subjects, while in the other two subjects (GIF and BMP) we could not run \textsc{AFLSmart}.
\Cref{tab:aflsmart} also includes the Vargha-Delaney effect sizes $A^{12}$ and the Wilcoxon signed-rank test for statistical significance.
We see that \textsc{AFL+FFMut} performed significantly better in four benchmarks, while \textsc{AFLSmart} had a significant advantage in two benchmarks.
As we again see in the differences, even on cases where \textsc{AFLSmart} performed better, \textsc{AFL+FFMut} has still covered some lines of code which were not covered by \textsc{AFLSmart}.

One contributing factor here is the completeness of the format specifications.
One advantage of \tool{} is that we can leverage very detailed binary templates which had already been developed for parsing.
The creators of \textsc{AFLSmart}, on the other hand, had to write their format specifications (\emph{Peach pits}) from scratch, so they are less complete.
For example, the \emph{Peach pit} for PNG only specializes the internal contents of three chunk types: IHDR, cHRM and IEND, while the PNG binary template fully defined 15 chunk types, even before any changes.

However, from our experience, the more detailed format specifications lead to only small improvements in fuzzing if those specifications are only used for parsing, as in \textsc{AFLSmart}.
\tool{} reaps the most benefits from those detailed specifications from its ability to also generate valid inputs, which is used in \textsc{FFMut} to apply semantically valid smart mutations which respect contextual information.

\takeaway{Using input formats for parsing, mutating, and generating, as \tool{} does, yields additional coverage over using them for parsing only, as AFLSmart does.}

We have also noticed that \tool{} is particularly efficient in generating high-quality inputs and achieving high coverage very quickly during the fuzzing campaign.

\begin{table}[h]
\rowcolors{2}{RawSienna!10}{white}
\caption{Line coverage (\%) of \tool{} and AFLSmart in \textbf{one hour} of fuzzing. Here, $A\setminus B$ represents the lines which were covered by technique $A$, but not $B$. Vargha-Delaney effect size $A^{12}$ (statistically significant effects in \textbf{bold}).}
\label{tab:aflsmart1}
\small
\begin{tabular}{lrrrrrrrrrr}
\rowcolor{RawSienna}
\evalhead \\
\textsc{AFLSmart} & 20.5 & 28.5 & --/-- & 10.4 & 8.3 & 33.9 & 17.1 & 8.0 & 22.1 & --/-- \\
\textsc{AFL+FFMut} & 27.3 & 31.4 & 71.9 & 10.4 & 8.5 & 37.2 & 18.1 & 8.2 & 22.2 & 30.0 \\
\hline
$\textsc{AFLSmart} \setminus \textsc{AFL+FFMut}$ & 0.0 & 0.4 & --/-- & 0.1 & 0.4 & 0.2 & 1.0 & 0.3 & 0.0 & --/-- \\
$\textsc{AFL+FFMut} \setminus \textsc{AFLSmart}$ & 6.8 & 3.3 & --/-- & 0.0 & 0.6 & 3.4 & 2.0 & 0.6 & 0.1 & --/-- \\
\hline
$A^{12}(\textsc{AFL+FFMut} > \textsc{AFLSmart})$ & \textbf{1.0} & \textbf{1.0} & --/-- & 0.43 & 0.73 & \textbf{1.0} & \textbf{0.92} & \textbf{0.86} & \textbf{0.97} & --/-- \\
\end{tabular}
\end{table}

\Cref{tab:aflsmart1} shows the same information as in \Cref{tab:aflsmart}, but only for the first hour of fuzzing (we use again an average over 10 runs).
We notice that, in the short run of one hour, \tool{} performed significantly better than \textsc{AFLSmart} on six out of eight benchmarks, and never performed significantly worse.

\takeaway{\tool{} is particularly good at finding coverage on a limited time budget.}

\subsection{RQ8: Bugs Found}

We close our evaluation with RQ8: \emph{\rqeight{}}
The answer is: Yes!
In our initial fuzzing experiments with \tool{}, we have found and reported:
\begin{itemize}
    \item 16 distinct segmentation faults (by different stack traces) and 8 distinct aborts in \texttt{ffmpeg} (MP4 and AVI), which turned out to be at least 8 distinct bugs which have already been fixed to date by the FFmpeg developers. These occur when \texttt{ffmpeg} is compiled with pthreads (which is the default option) and the fuzzer runs with a memory limit of 200~MB (using AFL's \texttt{-m} option).
These bugs are located in a wide range of different stages through the execution: allocation, initialization of h264 slices, multi-threading, writing \texttt{mov} packets, video encoding, decoding spectral data and context cleanup.
    \item 19 distinct memory errors (by different stack traces) in \texttt{timidity} (MIDI). We will update the paper when those bugs are fixed.
\end{itemize}
Note that \texttt{ffmpeg} ``is part of the workflow of hundreds of other software projects, and its libraries are a core part of software media players such as VLC, and has been included in core processing for YouTube and iTunes.''~\cite{wikipedia-ffmpeg}

\takeaway{\tool{} finds bugs in relevant software.}

\subsection{Discussion}
\label{sec:discussion}

Leveraging input format specifications pays off.
In every single setting, \tool{} explored additional coverage---and thus additional chances to find bugs and vulnerabilities.
\tool{} can be used as a standalone generator, notably in black-box settings; the integration with feedback-driven fuzzers such as AFL combines the best of format-aware and feedback-directed fuzzing.

The evaluation also shows the \emph{versatility} of \tool{}.
Its individual components (parsing/mutating/generating) can either be used standalone (say, \emph{generating} in a black-box setting),
or fully integrated with format-agnostic fuzzers (where the popular AFL tool can be replaced by any even more performant format-agnostic fuzzer).

This versatility and modularity makes \tool{} a \emph{platform} for quickly creating and integrating novel fuzzing strategies and adapting them towards the setting at hand---in contrast to fuzzers which have no knowledge about the input structure, or which use input specifications exclusively for parsing or for generating.
As our evaluation shows, both format-aware and format-agnostic strategies have their specific strengths, further motivating the need for an integrative platform.

Another important aspect not addressed previously is that, by construction, \tool{} gives the tester \emph{control} over what should be tested.
By commenting out particular parts of a binary template, \tool{} allows to focus on specific features---say, those that were recently changed or otherwise are critical.
This adds to the versatility of fuzzing with \tool{}.

\section{Related Work}
\label{sec:related}

\subsection{Fuzzing with Input Specifications}

Using language specifications for generating inputs is an old idea, especially
for context-free grammars. The different
grammars such as regular, context-free, context-sensitive, and
unconstrained grammars were invented by Chomsky for linguistic
applications~\cite{chomsky1956three} in 1950s specifically for parsing.
Using grammars for input generation was first suggested by
Burkhardt~\cite{burkhardt1967generating}, Hanford~\cite{hanford1970automatic}
and Purdom~\cite{purdom1972a} in the late 1960s and early 1970s.
Logic languages~\cite{green1970application}
such as Prolog~\cite{warren1977prolog} which are known from 1970s allow turning
the entire program inside out, using the program itself to generate the
kind of inputs that it accepts.
Free generators~\cite{GoldsteinOOPSLA2022} are a recent formalism which also unifies parsing with generation, as in \tool{}.
However, they focus on the generation of data structures, such as type-classes in Haskell, while \tool{} targets binary formats and integrates with existing format-agnostic fuzzers.

\subsection{Context-Free Grammars}

Grammar-based testing took off in the new millennium when researchers
rediscovered the utility of fuzzing, and how grammars can improve the
effectiveness of fuzzing. One of the first to recognize their use in
fuzzing was Godefroid~\cite{godefroid2008grammar} who augmented whitebox
fuzzing with grammars.
Other noteworthy grammar fuzzers include
Gramfuzz~\cite{guo2013gramfuzz}, Grammarinator~\cite{hodovan2018grammarinator},
Dharma~\cite{mozilla2019dharma}, Domato~\cite{gratric2019domato}, and CSS
Fuzz~\cite{ruderman2007introducing}, as well as PolyGlot~\cite{chen2021one}, which augments context-free grammars with semantic annotations.
LangFuzz~\cite{langfuzz} uses a language specification to collect code fragments which can be applied as smart mutations.
All these work on context-free languages, which are not sufficient to specify binary formats.

\subsection{Other Specification Languages}

Besides context-free grammars, fuzzers have used alternate input specifications, such as regular languages (i.e. finite state automata)~\cite{cui2014novel,wang2017automatic}, or constraint languages~\cite{dewey2014language}.
SISL~\cite{tempel2022sisl} uses a partial specification which is complemented by the use of concolic testing~\cite{cute}.
MoWF~\cite{pham2016model} leverages
a specification expressed as a constraint over the input space.
It uses selective symbolic execution to identify uncovered
branches, and can repair length fields, checksums and other validation fields.
The fuzzer by Pan et al.~\cite{pan2013efficient} leverages
higher order attribute grammars to describe length, checksums, and other
validation fields, and uses them to generate file format inputs such as PNGs.
Parsifal~\cite{levillain2014parsifal} targets both parsing and generation of binary formats. It is, however, limited to fixed size formats.
Nail~\cite{bangert2014nail} defines a parser generator
specification, and targets binary formats (typically protocols) that contain
offset fields and checksums, and unlike Parsifal, can handle more complex
constructs.
Underwood~\cite{underwood2012grammar} extends context-free grammars using
attribute grammars, and provides a mapping from binary format specifications
to attribute grammars. Such grammars can be used for parsing as well as
generation.
Beginner's~luck~\cite{lampropoulos2017beginner} is a language for generators that allows the integration of sampling constraints to be used during fuzzing.
libprotobuf-mutator~\cite{serebryany2017structure} uses protocol buffers to describe file formats, allowing the mutations to be done over the compact protocol buffer representation.
However, one disadvantage compared to \tool{} is the need to write a converter between the protocol buffer and the corresponding binary file.

In principle, \tool{} could make use of any of these format specifications, and still leverage its unique capabilities such as generating synchronized parsers and generators, or integrating format-agnostic fuzzers. By using binary templates, however, \tool{} can build on hundreds of format specifications that have been created and refined by an enthusiastic community for more than two decades now.

\subsection{Fuzzing Strategies}

CSmith~\cite{yang2011finding} shows that embedding the entire language
specification into the generator can lead to effective fuzzing. However, with
that, one loses the generality of being able to target other kinds of inputs.
Other notable research on grammar-based fuzzers include
LangFuzz~\cite{langfuzz}, Blendfuzz~\cite{yan2013structurized},
Skyfire~\cite{wang2017skyfire}.
\emph{Parameter sequences} in Zest~\cite{padhye2019semantic} and JQF~\cite{padhye2019jqf} encode generator choices like our \emph{decision seeds} and can be used for generator-based testing in the style of QuickCheck~\cite{quickcheck}.
Crowbar~\cite{crowbar} and CGPT~\cite{lampropoulos2019coverage} similarly also combine generator-based testing with coverage feedback, allowing both generation-based and mutation-based fuzzing.
However, obtaining the decision seeds by parsing existing files, as well as applying smart mutations on them (\Cref{sec:mutating}) is a novel contribution of \tool{}.

Several grammar-based fuzzers also incorporate coverage feedback from AFL.
Notable examples include Superion~\cite{wang2019superion}, Nautilus~\cite{aschermann2019nautilus} and Grimoire~\cite{blazytko2019grimoire}.
However, those fuzzers were only applied to text-based grammar input formats, such as markup languages (XML) or programming languages (JavaScript, PHP, Ruby, Lua, C, nasm, SQL, SMT).
So they would likely be unable to support features required for binary file formats, such as size fields, checksums, or bitfields.
Superion and Nautilus would require a complete format specification to be written from scratch for each new format.
Grimoire, on the other hand, can mine format specifications, but only for textual languages.

\subsection{Fuzzing Binary Formats}

WEIZZ~\cite{fioraldi2020weizz} and FFAFuzz~\cite{chen2022fast} are fuzzers which target chunk-based binary
formats. During fuzzing, they try to learn how the chunk-based format is specified.
Peach~\cite{eddington2011peach} is another fuzzer that targets binary formats.
AFL\-Smart~\cite{pham2019smart} is a fuzzer that provides chunk aware mutations
to the input, and hence targets binary formats. It does this by maintaining
a virtual structure of the input being fuzzed in memory.
AFLSmart is the closest related work to \tool{}.
However, since its format specifications (\emph{Peach pits}) had to be developed from scratch for use with AFLSmart, they are less complete and comprehensive than our binary templates.
Since \tool{} can use its binary templates not only for parsing, but also for generation, this enables additional fuzzing strategies detailed in this paper.

\section{Conclusion}
\label{sec:conclusion}

In order to fuzz binary inputs, the most powerful solution is to create a dedicated fuzzer.
However, creating such fuzzers takes significant effort.
With \tool{}, one can leverage \emph{hundreds of existing binary templates} to make existing fuzzers format-aware. \tool{} can readily use such binary templates as \emph{parsers,} which reveal the full structure of inputs.
Extending existing binary templates for generation unlocks fuzzing in black-box settings that are infeasible for format-agnostic fuzzers. By providing decision seeds and smart mutations, \tool{} can make any fuzzer format-aware.

While \tool{} is a highly efficient choice for domain-specific fuzzing, its modularity also makes it a useful platform for creating future fuzzers. Besides extending \tool{} with further formats, our future work will focus on the following topics:

\begin{description}
  \item[Black-box strategies.] Recent advances in grammar-based fuzzing, such as systematically achieving grammar coverage~\cite{havrikov2019ase} or learning and leveraging probability distributions~\cite{soremekun2020inputs} could easily be adopted for binary template fuzzing, too.
  \item[Search-based fuzzing.] The \emph{decision seed} format opens way for easily integrating alternate fuzzing strategies. Of particular interest is \emph{genetic optimization} as part of \emph{search-based testing,} since \tool{} already implements mutation and crossover operations.
  \item[Mining binary formats.] Recent techniques for mining context-free grammars from parsers \cite{gopinath2020mining,mathis2019parser} could be adapted to binary formats, simplifying template construction.
  \item[Mining binary constraints.] Extending a (mined) syntactic template,  we can track processing of individual input elements using dynamic tainting and dynamic analysis, and extract context-sensitive constraints from input processors.
  \item[Engaging the community.] We will be creating tutorials and other material to engage the community in writing binary templates. The Wikipedia list of file formats~\cite{wikipedia-files} lists more than 1,000 formats---so there is still lots to do!
\end{description}

\tool{} and all supporting material is available as open source. For more information on \tool{}, see its project page
\begin{center}
\toolURL
\end{center}

\section*{Acknowledgements}

We thank Pit Jost, Andreas Hanuja and Steven Dlucik for their work in supporting binary templates for multiple formats.

\bibliographystyle{ACM-Reference-Format}
\bibliography{FormatFuzzer}


\begin{thebibliography}{58}


\ifx \showCODEN    \undefined \def \showCODEN     #1{\unskip}     \fi
\ifx \showDOI      \undefined \def \showDOI       #1{#1}\fi
\ifx \showISBNx    \undefined \def \showISBNx     #1{\unskip}     \fi
\ifx \showISBNxiii \undefined \def \showISBNxiii  #1{\unskip}     \fi
\ifx \showISSN     \undefined \def \showISSN      #1{\unskip}     \fi
\ifx \showLCCN     \undefined \def \showLCCN      #1{\unskip}     \fi
\ifx \shownote     \undefined \def \shownote      #1{#1}          \fi
\ifx \showarticletitle \undefined \def \showarticletitle #1{#1}   \fi
\ifx \showURL      \undefined \def \showURL       {\relax}        \fi
\providecommand\bibfield[2]{#2}
\providecommand\bibinfo[2]{#2}
\providecommand\natexlab[1]{#1}
\providecommand\showeprint[2][]{arXiv:#2}

\bibitem[wik(2021a)]%
        {wikipedia-ffmpeg}
 \bibinfo{year}{2021}\natexlab{a}.
\newblock \bibinfo{title}{Wikipedia: ffmpeg}.
\newblock \bibinfo{howpublished}{\url{https://en.wikipedia.org/wiki/FFmpeg}}.
\newblock
\newblock
\shownote{Accessed May 2021}.


\bibitem[wik(2021b)]%
        {wikipedia-files}
 \bibinfo{year}{2021}\natexlab{b}.
\newblock \bibinfo{title}{Wikipedia: List of File Formats}.
\newblock
  \bibinfo{howpublished}{\url{https://en.wikipedia.org/wiki/List_of_file_formats}}.
\newblock
\newblock
\shownote{Accessed May 2021}.


\bibitem[Aschermann et~al\mbox{.}(2019)]%
        {aschermann2019nautilus}
\bibfield{author}{\bibinfo{person}{Cornelius Aschermann},
  \bibinfo{person}{Tommaso Frassetto}, \bibinfo{person}{Thorsten Holz},
  \bibinfo{person}{Patrick Jauernig}, \bibinfo{person}{Ahmad-Reza Sadeghi},
  {and} \bibinfo{person}{Daniel Teuchert}.} \bibinfo{year}{2019}\natexlab{}.
\newblock \showarticletitle{{NAUTILUS:} Fishing for Deep Bugs with Grammars}.
  In \bibinfo{booktitle}{\emph{Proceedings of NDSS 2019}}.
\newblock
\urldef\tempurl%
\url{https://www.ndss-symposium.org/ndss-paper/nautilus-fishing-for-deep-bugs-with-grammars/}
\showURL{%
\tempurl}


\bibitem[Bangert and Zeldovich(2014)]%
        {bangert2014nail}
\bibfield{author}{\bibinfo{person}{Julian Bangert} {and}
  \bibinfo{person}{Nickolai Zeldovich}.} \bibinfo{year}{2014}\natexlab{}.
\newblock \showarticletitle{Nail: A practical tool for parsing and generating
  data formats}. In \bibinfo{booktitle}{\emph{11th $\{$USENIX$\}$ Symposium on
  Operating Systems Design and Implementation ($\{$OSDI$\}$ 14)}}.
  \bibinfo{pages}{615--628}.
\newblock


\bibitem[Blazytko et~al\mbox{.}(2019)]%
        {blazytko2019grimoire}
\bibfield{author}{\bibinfo{person}{Tim Blazytko}, \bibinfo{person}{Matt
  Bishop}, \bibinfo{person}{Cornelius Aschermann}, \bibinfo{person}{Justin
  Cappos}, \bibinfo{person}{Moritz Schl{\"o}gel}, \bibinfo{person}{Nadia
  Korshun}, \bibinfo{person}{Ali Abbasi}, \bibinfo{person}{Marco
  Schweighauser}, \bibinfo{person}{Sebastian Schinzel}, \bibinfo{person}{Sergej
  Schumilo}, {et~al\mbox{.}}} \bibinfo{year}{2019}\natexlab{}.
\newblock \showarticletitle{$\{$GRIMOIRE$\}$: Synthesizing structure while
  fuzzing}. In \bibinfo{booktitle}{\emph{28th $\{$USENIX$\}$ Security Symposium
  ($\{$USENIX$\}$ Security 19)}}. \bibinfo{pages}{1985--2002}.
\newblock


\bibitem[Burkhardt(1967)]%
        {burkhardt1967generating}
\bibfield{author}{\bibinfo{person}{W.~H. Burkhardt}.}
  \bibinfo{year}{1967}\natexlab{}.
\newblock \showarticletitle{Generating test programs from syntax}.
\newblock \bibinfo{journal}{\emph{Computing}} \bibinfo{volume}{2},
  \bibinfo{number}{1} (\bibinfo{date}{March} \bibinfo{year}{1967}),
  \bibinfo{pages}{53--73}.
\newblock
\showISSN{1436-5057}
\urldef\tempurl%
\url{https://doi.org/10.1007/BF02235512}
\showDOI{\tempurl}


\bibitem[Chen et~al\mbox{.}(2021)]%
        {chen2021one}
\bibfield{author}{\bibinfo{person}{Yongheng Chen}, \bibinfo{person}{Rui Zhong},
  \bibinfo{person}{Hong Hu}, \bibinfo{person}{Hangfan Zhang},
  \bibinfo{person}{Yupeng Yang}, \bibinfo{person}{Dinghao Wu}, {and}
  \bibinfo{person}{Wenke Lee}.} \bibinfo{year}{2021}\natexlab{}.
\newblock \showarticletitle{{One Engine to Fuzz 'em All: Generic Language
  Processor Testing with Semantic Validation (to appear)}}. In
  \bibinfo{booktitle}{\emph{Proceedings of the 42nd IEEE Symposium on Security
  and Privacy (Oakland)}}. \bibinfo{address}{San Francisco, CA}.
\newblock


\bibitem[Chen et~al\mbox{.}(2022)]%
        {chen2022fast}
\bibfield{author}{\bibinfo{person}{Zehan Chen}, \bibinfo{person}{Yuliang Lu},
  \bibinfo{person}{Kailong Zhu}, \bibinfo{person}{Lu Yu}, {and}
  \bibinfo{person}{Jiazhen Zhao}.} \bibinfo{year}{2022}\natexlab{}.
\newblock \showarticletitle{Fast Format-Aware Fuzzing for Structured Input
  Applications}.
\newblock \bibinfo{journal}{\emph{Applied Sciences}} \bibinfo{volume}{12},
  \bibinfo{number}{18} (\bibinfo{year}{2022}), \bibinfo{pages}{9350}.
\newblock


\bibitem[Chomsky(1956)]%
        {chomsky1956three}
\bibfield{author}{\bibinfo{person}{Noam Chomsky}.}
  \bibinfo{year}{1956}\natexlab{}.
\newblock \showarticletitle{Three models for the description of language}.
\newblock \bibinfo{journal}{\emph{IRE Transactions on Information Theory}}
  \bibinfo{volume}{2} (\bibinfo{year}{1956}), \bibinfo{pages}{113--124}.
\newblock
\urldef\tempurl%
\url{https://chomsky.info/wp-content/uploads/195609-.pdf}
\showURL{%
\tempurl}


\bibitem[Claessen and Hughes(2011)]%
        {quickcheck}
\bibfield{author}{\bibinfo{person}{Koen Claessen} {and} \bibinfo{person}{John
  Hughes}.} \bibinfo{year}{2011}\natexlab{}.
\newblock \showarticletitle{QuickCheck: a lightweight tool for random testing
  of Haskell programs}.
\newblock \bibinfo{journal}{\emph{{ACM SIGPLAN} Notices}} \bibinfo{volume}{46},
  \bibinfo{number}{4} (\bibinfo{year}{2011}), \bibinfo{pages}{53--64}.
\newblock


\bibitem[Cui et~al\mbox{.}(2014)]%
        {cui2014novel}
\bibfield{author}{\bibinfo{person}{Baojiang Cui}, \bibinfo{person}{Shurui
  Liang}, \bibinfo{person}{Shilei Chen}, \bibinfo{person}{Bing Zhao}, {and}
  \bibinfo{person}{Xiaobing Liang}.} \bibinfo{year}{2014}\natexlab{}.
\newblock \showarticletitle{A novel fuzzing method for Zigbee based on finite
  state machine}.
\newblock \bibinfo{journal}{\emph{International Journal of Distributed Sensor
  Networks}} \bibinfo{volume}{10}, \bibinfo{number}{1} (\bibinfo{year}{2014}),
  \bibinfo{pages}{762891}.
\newblock


\bibitem["d0c\_s4vage" Johnson(2020a)]%
        {pfp}
\bibfield{author}{\bibinfo{person}{James "d0c\_s4vage" Johnson}.}
  \bibinfo{year}{2020}\natexlab{a}.
\newblock \bibinfo{title}{GitHub - d0c-s4vage/pfp: pfp - Python Format Parser -
  a python-based 010 Editor template interpreter}.
\newblock \bibinfo{howpublished}{\url{https://github.com/d0c-s4vage/pfp}}.
\newblock
\newblock
\shownote{Accessed August 1, 2021}.


\bibitem["d0c\_s4vage" Johnson(2020b)]%
        {py010parser}
\bibfield{author}{\bibinfo{person}{James "d0c\_s4vage" Johnson}.}
  \bibinfo{year}{2020}\natexlab{b}.
\newblock \bibinfo{title}{GitHub - d0c-s4vage/py010parser: A modified pycparser
  to parse 010 templates}.
\newblock
  \bibinfo{howpublished}{\url{https://github.com/d0c-s4vage/py010parser}}.
\newblock
\newblock
\shownote{Accessed August 1, 2021}.


\bibitem[David~Duce(2003)]%
        {png}
\bibfield{author}{\bibinfo{person}{Oxford Brookes University (Second~Edition)
  David~Duce}.} \bibinfo{year}{2003}\natexlab{}.
\newblock \bibinfo{title}{Portable Network Graphics (PNG) Specification (Second
  Edition)}.
\newblock \bibinfo{howpublished}{\url{https://www.w3.org/TR/PNG/}}.
\newblock
\newblock
\shownote{Accessed November 15, 2021}.


\bibitem[Dewey et~al\mbox{.}(2014)]%
        {dewey2014language}
\bibfield{author}{\bibinfo{person}{Kyle Dewey}, \bibinfo{person}{Jared Roesch},
  {and} \bibinfo{person}{Ben Hardekopf}.} \bibinfo{year}{2014}\natexlab{}.
\newblock \showarticletitle{Language fuzzing using constraint logic
  programming}. ACM, \bibinfo{pages}{725--730}.
\newblock


\bibitem[Dolan(2021)]%
        {crowbar}
\bibfield{author}{\bibinfo{person}{Stephen Dolan}.}
  \bibinfo{year}{2021}\natexlab{}.
\newblock \bibinfo{title}{Crowbar}.
\newblock \bibinfo{howpublished}{\url{https://github.com/stedolan/crowbar}}.
\newblock
\newblock
\shownote{Accessed November 15, 2021}.


\bibitem[Eddington(2011)]%
        {eddington2011peach}
\bibfield{author}{\bibinfo{person}{Michael Eddington}.}
  \bibinfo{year}{2011}\natexlab{}.
\newblock \showarticletitle{Peach fuzzing platform}.
\newblock \bibinfo{journal}{\emph{Peach Fuzzer}}  \bibinfo{volume}{34}
  (\bibinfo{year}{2011}).
\newblock


\bibitem[Fioraldi et~al\mbox{.}(2020a)]%
        {fioraldi2020weizz}
\bibfield{author}{\bibinfo{person}{Andrea Fioraldi},
  \bibinfo{person}{Daniele~Cono D'Elia}, {and} \bibinfo{person}{Emilio Coppa}.}
  \bibinfo{year}{2020}\natexlab{a}.
\newblock \showarticletitle{WEIZZ: Automatic grey-box fuzzing for structured
  binary formats}. In \bibinfo{booktitle}{\emph{Proceedings of the 29th ACM
  SIGSOFT International Symposium on Software Testing and Analysis}}.
  \bibinfo{pages}{1--13}.
\newblock


\bibitem[Fioraldi et~al\mbox{.}(2020b)]%
        {AFLplusplus-Woot20}
\bibfield{author}{\bibinfo{person}{Andrea Fioraldi}, \bibinfo{person}{Dominik
  Maier}, \bibinfo{person}{Heiko Ei{\ss}feldt}, {and} \bibinfo{person}{Marc
  Heuse}.} \bibinfo{year}{2020}\natexlab{b}.
\newblock \showarticletitle{{AFL++}: Combining Incremental Steps of Fuzzing
  Research}. In \bibinfo{booktitle}{\emph{14th {USENIX} Workshop on Offensive
  Technologies ({WOOT} 20)}}. \bibinfo{publisher}{{USENIX} Association}.
\newblock


\bibitem[Fratric(2019)]%
        {gratric2019domato}
\bibfield{author}{\bibinfo{person}{Ivan Fratric}.}
  \bibinfo{year}{2019}\natexlab{}.
\newblock \bibinfo{booktitle}{\emph{{Domato A DOM fuzzer}}}.
\newblock
\urldef\tempurl%
\url{https://github.com/googleprojectzero/domato}
\showURL{%
\tempurl}


\bibitem[Godefroid et~al\mbox{.}(2008)]%
        {godefroid2008grammar}
\bibfield{author}{\bibinfo{person}{Patrice Godefroid}, \bibinfo{person}{Adam
  Kiezun}, {and} \bibinfo{person}{Michael~Y. Levin}.}
  \bibinfo{year}{2008}\natexlab{}.
\newblock \showarticletitle{Grammar-based Whitebox Fuzzing}.
  \bibinfo{publisher}{ACM}, \bibinfo{address}{New York, NY, USA},
  \bibinfo{pages}{206--215}.
\newblock
\showISBNx{978-1-59593-860-2}


\bibitem[Goldstein and Pierce(2022)]%
        {GoldsteinOOPSLA2022}
\bibfield{author}{\bibinfo{person}{Harrison Goldstein} {and}
  \bibinfo{person}{Benjamin~C. Pierce}.} \bibinfo{year}{2022}\natexlab{}.
\newblock \showarticletitle{Parsing Randomness}.
\newblock \bibinfo{journal}{\emph{Proc. ACM Program. Lang.}}
  \bibinfo{number}{OOPSLA} (\bibinfo{year}{2022}).
\newblock


\bibitem[Gopinath et~al\mbox{.}(2020)]%
        {gopinath2020mining}
\bibfield{author}{\bibinfo{person}{Rahul Gopinath}, \bibinfo{person}{Bj{\"o}rn
  Mathis}, {and} \bibinfo{person}{Andreas Zeller}.}
  \bibinfo{year}{2020}\natexlab{}.
\newblock \showarticletitle{Mining input grammars from dynamic control flow}.
  In \bibinfo{booktitle}{\emph{Proceedings of the 28th ACM Joint Meeting on
  European Software Engineering Conference and Symposium on the Foundations of
  Software Engineering}}. \bibinfo{pages}{172--183}.
\newblock


\bibitem[Green(1970)]%
        {green1970application}
\bibfield{author}{\bibinfo{person}{Claude~Cordell Green}.}
  \bibinfo{year}{1970}\natexlab{}.
\newblock \bibinfo{booktitle}{\emph{The application of theorem proving to
  question-answering systems}}.
\newblock Number~96. \bibinfo{publisher}{Management Information Services}.
\newblock


\bibitem[Guo et~al\mbox{.}(2013)]%
        {guo2013gramfuzz}
\bibfield{author}{\bibinfo{person}{Tao Guo}, \bibinfo{person}{Puhan Zhang},
  \bibinfo{person}{Xin Wang}, {and} \bibinfo{person}{Qiang Wei}.}
  \bibinfo{year}{2013}\natexlab{}.
\newblock \showarticletitle{Gramfuzz: Fuzzing testing of web browsers based on
  grammar analysis and structural mutation}. In \bibinfo{booktitle}{\emph{2013
  Second International Conference on Informatics \& Applications (ICIA)}}.
  IEEE, \bibinfo{pages}{212--215}.
\newblock


\bibitem[Hanford(1970)]%
        {hanford1970automatic}
\bibfield{author}{\bibinfo{person}{Kenneth~V. Hanford}.}
  \bibinfo{year}{1970}\natexlab{}.
\newblock \showarticletitle{Automatic Generation of Test Cases}.
\newblock \bibinfo{journal}{\emph{IBM Syst. J.}} \bibinfo{volume}{9},
  \bibinfo{number}{4} (\bibinfo{date}{Dec.} \bibinfo{year}{1970}),
  \bibinfo{pages}{242--257}.
\newblock
\showISSN{0018-8670}
\urldef\tempurl%
\url{https://doi.org/10.1147/sj.94.0242}
\showDOI{\tempurl}


\bibitem[Havrikov and Zeller(2019)]%
        {havrikov2019ase}
\bibfield{author}{\bibinfo{person}{Nikolas Havrikov} {and}
  \bibinfo{person}{Andreas Zeller}.} \bibinfo{year}{2019}\natexlab{}.
\newblock \showarticletitle{Systematically Covering Input Structure}. In
  \bibinfo{booktitle}{\emph{Proceedings of the 34th IEEE/ACM International
  Conference on Automated Software Engineering}} (San Diego, California)
  \emph{(\bibinfo{series}{ASE '19})}. \bibinfo{publisher}{IEEE Press},
  \bibinfo{pages}{189–199}.
\newblock
\showISBNx{9781728125084}
\urldef\tempurl%
\url{https://doi.org/10.1109/ASE.2019.00027}
\showDOI{\tempurl}


\bibitem[Hodov{\'a}n et~al\mbox{.}(2018)]%
        {hodovan2018grammarinator}
\bibfield{author}{\bibinfo{person}{Ren{\'a}ta Hodov{\'a}n},
  \bibinfo{person}{{\'A}kos Kiss}, {and} \bibinfo{person}{Tibor Gyim{\'o}thy}.}
  \bibinfo{year}{2018}\natexlab{}.
\newblock \showarticletitle{Grammarinator: a grammar-based open source fuzzer}.
  In \bibinfo{booktitle}{\emph{Proceedings of the 9th ACM SIGSOFT International
  Workshop on Automating TEST Case Design, Selection, and Evaluation}}. ACM,
  \bibinfo{pages}{45--48}.
\newblock


\bibitem[Holler et~al\mbox{.}(2012)]%
        {langfuzz}
\bibfield{author}{\bibinfo{person}{Christian Holler}, \bibinfo{person}{Kim
  Herzig}, {and} \bibinfo{person}{Andreas Zeller}.}
  \bibinfo{year}{2012}\natexlab{}.
\newblock \showarticletitle{Fuzzing with Code Fragments}. In
  \bibinfo{booktitle}{\emph{Proceedings of the 21st USENIX Conference on
  Security Symposium}} (Bellevue, WA) \emph{(\bibinfo{series}{Security'12})}.
  \bibinfo{publisher}{USENIX Association}, \bibinfo{address}{Berkeley, CA,
  USA}, \bibinfo{pages}{38--38}.
\newblock


\bibitem[Lampropoulos et~al\mbox{.}(2017)]%
        {lampropoulos2017beginner}
\bibfield{author}{\bibinfo{person}{Leonidas Lampropoulos},
  \bibinfo{person}{Diane Gallois-Wong}, \bibinfo{person}{C{\u{a}}t{\u{a}}lin
  Hri{\c{t}}cu}, \bibinfo{person}{John Hughes}, \bibinfo{person}{Benjamin~C
  Pierce}, {and} \bibinfo{person}{Li-yao Xia}.}
  \bibinfo{year}{2017}\natexlab{}.
\newblock \showarticletitle{Beginner's luck: a language for property-based
  generators}. In \bibinfo{booktitle}{\emph{Proceedings of the 44th ACM SIGPLAN
  Symposium on Principles of Programming Languages}}.
  \bibinfo{pages}{114--129}.
\newblock


\bibitem[Lampropoulos et~al\mbox{.}(2019)]%
        {lampropoulos2019coverage}
\bibfield{author}{\bibinfo{person}{Leonidas Lampropoulos},
  \bibinfo{person}{Michael Hicks}, {and} \bibinfo{person}{Benjamin~C Pierce}.}
  \bibinfo{year}{2019}\natexlab{}.
\newblock \showarticletitle{Coverage guided, property based testing}.
\newblock \bibinfo{journal}{\emph{Proceedings of the ACM on Programming
  Languages}} \bibinfo{volume}{3}, \bibinfo{number}{OOPSLA}
  (\bibinfo{year}{2019}), \bibinfo{pages}{1--29}.
\newblock


\bibitem[Levillain(2014)]%
        {levillain2014parsifal}
\bibfield{author}{\bibinfo{person}{Olivier Levillain}.}
  \bibinfo{year}{2014}\natexlab{}.
\newblock \showarticletitle{Parsifal: A pragmatic solution to the binary
  parsing problems}. In \bibinfo{booktitle}{\emph{2014 IEEE Security and
  Privacy Workshops}}. IEEE, \bibinfo{pages}{191--197}.
\newblock


\bibitem[Mathis et~al\mbox{.}(2019)]%
        {mathis2019parser}
\bibfield{author}{\bibinfo{person}{Bj{\"o}rn Mathis}, \bibinfo{person}{Rahul
  Gopinath}, \bibinfo{person}{Micha{\"e}l Mera}, \bibinfo{person}{Alexander
  Kampmann}, \bibinfo{person}{Matthias H{\"o}schele}, {and}
  \bibinfo{person}{Andreas Zeller}.} \bibinfo{year}{2019}\natexlab{}.
\newblock \showarticletitle{Parser-directed fuzzing}. In
  \bibinfo{booktitle}{\emph{Proceedings of the 40th ACM SIGPLAN Conference on
  Programming Language Design and Implementation}}. \bibinfo{pages}{548--560}.
\newblock


\bibitem[Mozilla(2019)]%
        {mozilla2019dharma}
\bibfield{author}{\bibinfo{person}{Mozilla}.} \bibinfo{year}{2019}\natexlab{}.
\newblock \bibinfo{booktitle}{\emph{{Dharma: A generation-based, context-free
  grammar fuzzer}}}.
\newblock
\urldef\tempurl%
\url{https://blog.mozilla.org/security/2015/06/29/dharma/}
\showURL{%
\tempurl}


\bibitem[Padhye et~al\mbox{.}(2019a)]%
        {padhye2019jqf}
\bibfield{author}{\bibinfo{person}{Rohan Padhye}, \bibinfo{person}{Caroline
  Lemieux}, {and} \bibinfo{person}{Koushik Sen}.}
  \bibinfo{year}{2019}\natexlab{a}.
\newblock \showarticletitle{JQF: coverage-guided property-based testing in
  Java}. In \bibinfo{booktitle}{\emph{Proceedings of the 28th ACM SIGSOFT
  International Symposium on Software Testing and Analysis}}.
  \bibinfo{pages}{398--401}.
\newblock


\bibitem[Padhye et~al\mbox{.}(2019b)]%
        {padhye2019semantic}
\bibfield{author}{\bibinfo{person}{Rohan Padhye}, \bibinfo{person}{Caroline
  Lemieux}, \bibinfo{person}{Koushik Sen}, \bibinfo{person}{Mike Papadakis},
  {and} \bibinfo{person}{Yves Le~Traon}.} \bibinfo{year}{2019}\natexlab{b}.
\newblock \showarticletitle{Semantic fuzzing with zest}. In
  \bibinfo{booktitle}{\emph{Proceedings of the 28th ACM SIGSOFT International
  Symposium on Software Testing and Analysis}}. \bibinfo{pages}{329--340}.
\newblock


\bibitem[Pan et~al\mbox{.}(2013)]%
        {pan2013efficient}
\bibfield{author}{\bibinfo{person}{Fan Pan}, \bibinfo{person}{Ying Hou},
  \bibinfo{person}{Zheng Hong}, \bibinfo{person}{Lifa Wu}, {and}
  \bibinfo{person}{Haiguang Lai}.} \bibinfo{year}{2013}\natexlab{}.
\newblock \showarticletitle{Efficient Model-based Fuzz Testing Using
  Higher-order Attribute Grammars.}
\newblock \bibinfo{journal}{\emph{JSW}} \bibinfo{volume}{8},
  \bibinfo{number}{3} (\bibinfo{year}{2013}), \bibinfo{pages}{645--651}.
\newblock


\bibitem[Pham et~al\mbox{.}(2016)]%
        {pham2016model}
\bibfield{author}{\bibinfo{person}{Van-Thuan Pham}, \bibinfo{person}{Marcel
  B{\"o}hme}, {and} \bibinfo{person}{Abhik Roychoudhury}.}
  \bibinfo{year}{2016}\natexlab{}.
\newblock \showarticletitle{Model-based whitebox fuzzing for program binaries}.
  In \bibinfo{booktitle}{\emph{Proceedings of the 31st IEEE/ACM International
  Conference on Automated Software Engineering}}. \bibinfo{pages}{543--553}.
\newblock


\bibitem[Pham et~al\mbox{.}(2019)]%
        {pham2019smart}
\bibfield{author}{\bibinfo{person}{Van-Thuan Pham}, \bibinfo{person}{Marcel
  B{\"o}hme}, \bibinfo{person}{Andrew~Edward Santosa},
  \bibinfo{person}{Alexandru~Razvan Caciulescu}, {and} \bibinfo{person}{Abhik
  Roychoudhury}.} \bibinfo{year}{2019}\natexlab{}.
\newblock \showarticletitle{Smart greybox fuzzing}.
\newblock \bibinfo{journal}{\emph{IEEE Transactions on Software Engineering}}
  (\bibinfo{year}{2019}).
\newblock


\bibitem[Purdom(1972)]%
        {purdom1972a}
\bibfield{author}{\bibinfo{person}{Paul Purdom}.}
  \bibinfo{year}{1972}\natexlab{}.
\newblock \showarticletitle{A sentence generator for testing parsers}.
\newblock \bibinfo{journal}{\emph{BIT Numerical Mathematics}}
  \bibinfo{volume}{12}, \bibinfo{number}{3} (\bibinfo{year}{1972}),
  \bibinfo{pages}{366--375}.
\newblock
\showISSN{0006-3835}
\urldef\tempurl%
\url{https://doi.org/10.1007/BF01932308}
\showDOI{\tempurl}


\bibitem[Ruderman(2007)]%
        {ruderman2007introducing}
\bibfield{author}{\bibinfo{person}{Jesse Ruderman}.}
  \bibinfo{year}{2007}\natexlab{}.
\newblock \bibinfo{booktitle}{\emph{Introducing jsfunfuzz}}.
\newblock
\urldef\tempurl%
\url{http://www.squarefree.com/2007/08/02/introducing-jsfunfuzz/}
\showURL{%
\tempurl}


\bibitem[Sen et~al\mbox{.}(2005)]%
        {cute}
\bibfield{author}{\bibinfo{person}{Koushik Sen}, \bibinfo{person}{Darko
  Marinov}, {and} \bibinfo{person}{Gul Agha}.} \bibinfo{year}{2005}\natexlab{}.
\newblock \showarticletitle{{CUTE}: A Concolic Unit Testing Engine for {C}}. In
  \bibinfo{booktitle}{\emph{ESEC/FSE'05}}.
\newblock


\bibitem[Serebryany(2016)]%
        {serebryany2016continuous}
\bibfield{author}{\bibinfo{person}{Kosta Serebryany}.}
  \bibinfo{year}{2016}\natexlab{}.
\newblock \showarticletitle{Continuous fuzzing with libfuzzer and
  addresssanitizer}. In \bibinfo{booktitle}{\emph{2016 IEEE Cybersecurity
  Development (SecDev)}}. IEEE, \bibinfo{pages}{157--157}.
\newblock


\bibitem[Serebryany et~al\mbox{.}(2017)]%
        {serebryany2017structure}
\bibfield{author}{\bibinfo{person}{Kostya Serebryany}, \bibinfo{person}{Vitaly
  Buka}, {and} \bibinfo{person}{Matt Morehouse}.}
  \bibinfo{year}{2017}\natexlab{}.
\newblock \showarticletitle{Structure-aware fuzzing for Clang and LLVM with
  libprotobuf-mutator}.
\newblock  (\bibinfo{year}{2017}).
\newblock


\bibitem[Software(2021a)]%
        {010editortemplatesrepository}
\bibfield{author}{\bibinfo{person}{SweetScape Software}.}
  \bibinfo{year}{2021}\natexlab{a}.
\newblock \bibinfo{title}{010 Editor - Binary Template Repository - Download
  Binary Templates}.
\newblock
  \bibinfo{howpublished}{\url{https://www.sweetscape.com/010editor/repository/templates/}}.
\newblock
\newblock
\shownote{Accessed August 1, 2021}.


\bibitem[Software(2021b)]%
        {010editortemplates}
\bibfield{author}{\bibinfo{person}{SweetScape Software}.}
  \bibinfo{year}{2021}\natexlab{b}.
\newblock \bibinfo{title}{010 Editor - Binary Templates - Parsing Binary
  Files}.
\newblock
  \bibinfo{howpublished}{\url{https://www.sweetscape.com/010editor/templates.html}}.
\newblock
\newblock
\shownote{Accessed August 1, 2021}.


\bibitem[Software(2021c)]%
        {010editor}
\bibfield{author}{\bibinfo{person}{SweetScape Software}.}
  \bibinfo{year}{2021}\natexlab{c}.
\newblock \bibinfo{title}{{010 Editor - Pro Text/Hex Editor | Edit 160+ Formats
  | Fast \& Powerful}}.
\newblock \bibinfo{howpublished}{\url{https://www.sweetscape.com/010editor/}}.
\newblock
\newblock
\shownote{Accessed August 1, 2021}.


\bibitem[Software(2021d)]%
        {010editortemplatelanguage}
\bibfield{author}{\bibinfo{person}{SweetScape Software}.}
  \bibinfo{year}{2021}\natexlab{d}.
\newblock \bibinfo{title}{010 Editor Manual - Writing Templates}.
\newblock
  \bibinfo{howpublished}{\url{https://www.sweetscape.com/010editor/manual/IntroTemplates.htm}}.
\newblock
\newblock
\shownote{Accessed August 1, 2021}.


\bibitem[{Soremekun} et~al\mbox{.}(2020)]%
        {soremekun2020inputs}
\bibfield{author}{\bibinfo{person}{Ezekiel {Soremekun}},
  \bibinfo{person}{Esteban {Pavese}}, \bibinfo{person}{Nikolas {Havrikov}},
  \bibinfo{person}{Lars {Grunske}}, {and} \bibinfo{person}{Andreas {Zeller}}.}
  \bibinfo{year}{2020}\natexlab{}.
\newblock \showarticletitle{Inputs from Hell: Learning Input Distributions for
  Grammar-Based Test Generation}.
\newblock \bibinfo{journal}{\emph{IEEE Transactions on Software Engineering}}
  (\bibinfo{year}{2020}), \bibinfo{pages}{1--1}.
\newblock
\urldef\tempurl%
\url{https://doi.org/10.1109/TSE.2020.3013716}
\showDOI{\tempurl}


\bibitem[Tempel et~al\mbox{.}(2022)]%
        {tempel2022sisl}
\bibfield{author}{\bibinfo{person}{S{\"o}ren Tempel}, \bibinfo{person}{Vladimir
  Herdt}, {and} \bibinfo{person}{Rolf Drechsler}.}
  \bibinfo{year}{2022}\natexlab{}.
\newblock \showarticletitle{SISL: Concolic Testing of Structured Binary Input
  Formats via Partial Specification}. In \bibinfo{booktitle}{\emph{Automated
  Technology for Verification and Analysis: 20th International Symposium, ATVA
  2022, Virtual Event, October 25--28, 2022, Proceedings}}. Springer,
  \bibinfo{pages}{77--82}.
\newblock


\bibitem[Underwood(2012)]%
        {underwood2012grammar}
\bibfield{author}{\bibinfo{person}{William Underwood}.}
  \bibinfo{year}{2012}\natexlab{}.
\newblock \showarticletitle{Grammar-Based Specification and Parsing of Binary
  File Formats}.
\newblock \bibinfo{journal}{\emph{International Journal of Digital Curation}}
  \bibinfo{volume}{7} (\bibinfo{date}{03} \bibinfo{year}{2012}),
  \bibinfo{pages}{95--106}.
\newblock
\urldef\tempurl%
\url{https://doi.org/10.2218/ijdc.v7i1.217}
\showDOI{\tempurl}


\bibitem[Wang et~al\mbox{.}(2017a)]%
        {wang2017skyfire}
\bibfield{author}{\bibinfo{person}{Junjie Wang}, \bibinfo{person}{Bihuan Chen},
  \bibinfo{person}{Lei Wei}, {and} \bibinfo{person}{Yang Liu}.}
  \bibinfo{year}{2017}\natexlab{a}.
\newblock \showarticletitle{Skyfire: Data-driven seed generation for fuzzing}.
  IEEE, \bibinfo{pages}{579--594}.
\newblock


\bibitem[Wang et~al\mbox{.}(2019)]%
        {wang2019superion}
\bibfield{author}{\bibinfo{person}{Junjie Wang}, \bibinfo{person}{Bihuan Chen},
  \bibinfo{person}{Lei Wei}, {and} \bibinfo{person}{Yang Liu}.}
  \bibinfo{year}{2019}\natexlab{}.
\newblock \showarticletitle{Superion: grammar-aware greybox fuzzing}. In
  \bibinfo{booktitle}{\emph{Proceedings of the 41st International Conference on
  Software Engineering}}. IEEE Press, \bibinfo{pages}{724--735}.
\newblock


\bibitem[Wang et~al\mbox{.}(2017b)]%
        {wang2017automatic}
\bibfield{author}{\bibinfo{person}{Ming-Hung Wang}, \bibinfo{person}{Han-Chi
  Wang}, \bibinfo{person}{You-Ru Chen}, {and} \bibinfo{person}{Chin-Laung
  Lei}.} \bibinfo{year}{2017}\natexlab{b}.
\newblock \showarticletitle{Automatic Test Pattern Generator for Fuzzing Based
  on Finite State Machine}.
\newblock \bibinfo{journal}{\emph{Security and Communication Networks}}
  \bibinfo{volume}{2017} (\bibinfo{year}{2017}).
\newblock


\bibitem[Warren et~al\mbox{.}(1977)]%
        {warren1977prolog}
\bibfield{author}{\bibinfo{person}{David~HD Warren}, \bibinfo{person}{Luis~M
  Pereira}, {and} \bibinfo{person}{Fernando Pereira}.}
  \bibinfo{year}{1977}\natexlab{}.
\newblock \showarticletitle{Prolog-the language and its implementation compared
  with Lisp}.
\newblock \bibinfo{journal}{\emph{ACM SIGPLAN Notices}} \bibinfo{volume}{12},
  \bibinfo{number}{8} (\bibinfo{year}{1977}), \bibinfo{pages}{109--115}.
\newblock


\bibitem[Yan et~al\mbox{.}(2013)]%
        {yan2013structurized}
\bibfield{author}{\bibinfo{person}{Jingbo Yan}, \bibinfo{person}{Yuqing Zhang},
  {and} \bibinfo{person}{Dingning Yang}.} \bibinfo{year}{2013}\natexlab{}.
\newblock \showarticletitle{Structurized grammar-based fuzz testing for
  programs with highly structured inputs}.
\newblock \bibinfo{journal}{\emph{Security and Communication Networks}}
  \bibinfo{volume}{6}, \bibinfo{number}{11} (\bibinfo{year}{2013}),
  \bibinfo{pages}{1319--1330}.
\newblock


\bibitem[Yang et~al\mbox{.}(2011)]%
        {yang2011finding}
\bibfield{author}{\bibinfo{person}{Xuejun Yang}, \bibinfo{person}{Yang Chen},
  \bibinfo{person}{Eric Eide}, {and} \bibinfo{person}{John Regehr}.}
  \bibinfo{year}{2011}\natexlab{}.
\newblock \showarticletitle{Finding and understanding bugs in C compilers}. In
  \bibinfo{booktitle}{\emph{ACM SIGPLAN Notices}}, Vol.~\bibinfo{volume}{46}.
  ACM, \bibinfo{pages}{283--294}.
\newblock


\bibitem[Zalewski(2016)]%
        {afl}
\bibfield{author}{\bibinfo{person}{Michał Zalewski}.}
  \bibinfo{year}{2016}\natexlab{}.
\newblock \bibinfo{title}{American Fuzzy Lop}.
\newblock \bibinfo{howpublished}{\url{http://lcamtuf.coredump.cx/afl}}.
\newblock
\newblock
\shownote{Accessed October 1, 2016}.


\end{thebibliography}

\end{document}